\documentclass[useAMS,usenatbib]{mn2e}
\usepackage{xcolor}
\usepackage{amsmath}
\usepackage{float}
\usepackage{txfonts}
\usepackage{fleqn}   
\usepackage{graphicx}
\usepackage{epstopdf}
\usepackage{verbatim}
\usepackage{mathtools}
\usepackage{bm}

\title[BSPWNe: emission and polarization]
{Full-3D relativistic MHD simulations of  bow shock pulsar wind nebulae: emission and polarization}
\author[B. Olmi \& N. Bucciantini]{
 B. Olmi$^{1,2,3}$ \thanks{E-mail: barbara@arcetri.astro.it} \& N. Bucciantini$^{1,4,5}$\\
$^{1}$INAF - Osservatorio Astrofisico di Arcetri, Largo E. Fermi 5,
I-50125 Firenze, Italy\\
$^{2}$Institute of Space Sciences (ICE, CSIC), Campus UAB, Carrer de Magrans s/n, 08193 Barcelona, Spain\\
$^{3}$Institut d'Estudis Espacials de Catalunya (IEEC), 08034 Barcelona, Spain\\
$^{4}$Dipartimento di Fisica e Astronomia, Universit\`a degli Studi di Firenze, Via G. Sansone 1, 
I-50019 Sesto F.~no  (Firenze), Italy\\
$^{5}$INFN - Sezione di Firenze, Via G. Sansone 1, I-50019 Sesto F.~no  (Firenze), Italy\\
}

\begin{document}
 
\date{Accepted / Received}

\maketitle

\label{firstpage}

\begin{abstract}
Bow shock pulsar wind nebulae are observed with a variety of complex morphologies at different wavelengths, most likely due to differences in the magnetic field strength and pulsar wind geometry. Here we present a detailed analysis, showing how these differences affect the observational properties in these systems, focusing on non-thermal synchrotron emission. By adopting different prescriptions for the local emissivity, on top of the magnetic and flow patterns taken from 3D high-resolution numerical simulations in relativistic MHD, and considering various viewing angles, we try to characterize the main features of the emission and polarization, to verify if and how these can be used to get information, or to put constraints, on known objects.
We found for example that conditions leading to a strong development of the turbulence in the bow shock tail produce substantial differences in the emission pattern, especially in polarized light.
\end{abstract}

\begin{keywords}
polarization - relativistic processes - ISM: supernova remnants - pulsars: general - methods: numerical -  radiation mechanisms: non-thermal
\end{keywords}

\section{Introduction}
\label{sec:intro}
Bow shock pulsar wind nebulae (BSPWNe) are a peculiar subclass of the larger set of pulsar wind nebulae (PWNe), defined broadly as non-thermal synchrotron emitting sources powered by the spin-down luminosity of a pulsar \citep{Gaensler:2006}. Unlike systems where the pulsar is still confined in the parent supernova remnant (SNR), and where the PWN is observed with a center-filled morphology, BSPWNe involve older objects,  where the pulsar is directly interacting with the interstellar medium (in a few cases BSPWNe are seen also for
pulsars interacting with the SNR shell). It is in fact estimated that a consistent fraction, between 10\% and 50\%, of all the pulsars is born with a kick velocity in the range 100-500 km s$^{-1}$ \citep{Cordes_Chernoff98a,Arzoumanian:2002,Sartore_Ripamonti+10a,Verbunt_Igoshev+17a}, while the progenitor remnant is in decelerated expansion \citep{Truelove_McKee99a,Cioffi_McKee+88a,Leahy_Green+14a,Sanchez-Cruces_Rosado+18a}.
They are thus fated to leave their parent SNR on short timescales if compared with the typical pulsar ages ($\sim 10^3$ vs $\sim 10^6$ years). 

Given that the typical sound speed in the interstellar medium (ISM) is of order 10-100 km s$^{-1}$, well below the typical pulsar velocities, as soon as the star leaves the SNR shell and starts to interact directly with the ISM, its motion becomes strongly supersonic. 
The balance of the pulsar wind ram pressure with the ram pressure of the surrounding ISM through which it moves, induces the formation of a bow shock
\citep{Wilkin:1996,Bucciantini:2001,Bucciantini:2002}, characterized by an elongated cometary morphology, with the pulsar located at the head
of a long tail of plasma, extending in the direction opposite to its motion.  
As in the case of other PWNe, the relativistic pulsar wind shocked and decelerated into a strong termination shock, inflates within this
cometary nebula a bubble of relativistic particles and magnetic fields, that is a synchrotron emitter, form radio to X-rays.  
Indeed many of such systems have been identified in recent years as non-thermal sources \citep{Arzoumanian_Cordes+04a,Kargaltsev:2017,Kargaltsev_Misanovic+08a,Gaensler:2004,Yusef-Zadeh:2005,Li:2005,Gaensler05a,Chatterjee:2005,Ng_Camilo+09a,Hales_Gaensler+09a,Ng_Gaensler+10a,De-Luca_Marelli+11a,Marelli_De-Luca+13a,Jakobsen_Tomsick+14a,Misanovic_Pavlov+08a,Posselt_Pavlov+17a,Klingler_Rangelov+16a,Ng:2012}. 
For a few of these objects radio polarimetric data are also available, pointing to a variety of magnetic configurations \citep{Ng:2012, Yusef-Zadeh:2005, Ng_Gaensler+10a, Klingler:2016, Kargaltsev:2017}.

In the case of pulsars moving in an ISM with neutral hydrogen these nebulae can be observed in H$_\alpha$ emission
\citep{Kulkarni_Hester88a,Cordes:93,Bell_Bailes+95a,van-Kerkwijk_Kulkarni01a,Jones_Stappers+02a,Brownsberger:2014,Romani_Slane+17a}, due to charge exchange and collisional excitation processed with the shocked ISM material \citep{Chevalier_Kirshner+80a,Hester_Raymond+94a,Bucciantini:2001,Ghavamian_Raymond+01a}, or alternatively  in the UV \citep{Rangelov_Pavlov+16a} and IR
\citep{Wang_Kaplan+13a}. 
It is debatable if many extended and morphologically complex TeV sources detected by HESS in coincidence with pulsars, can be attributed to the BSPWN class \citep{Kargaltsev:2012}, especially in those cases where the PSR is strongly offsetted from the center of emission. 
An example of this kind of uncertainty is represented by the young and energetic PSR J05376-6910 from 
the Large Magellanic Cloud, which is not uniquely identified as a bow shock nebula due to the large distance \citep{Kargaltsev:2017}.
\begin{figure}
	\centering
	\includegraphics[width=.5\textwidth]{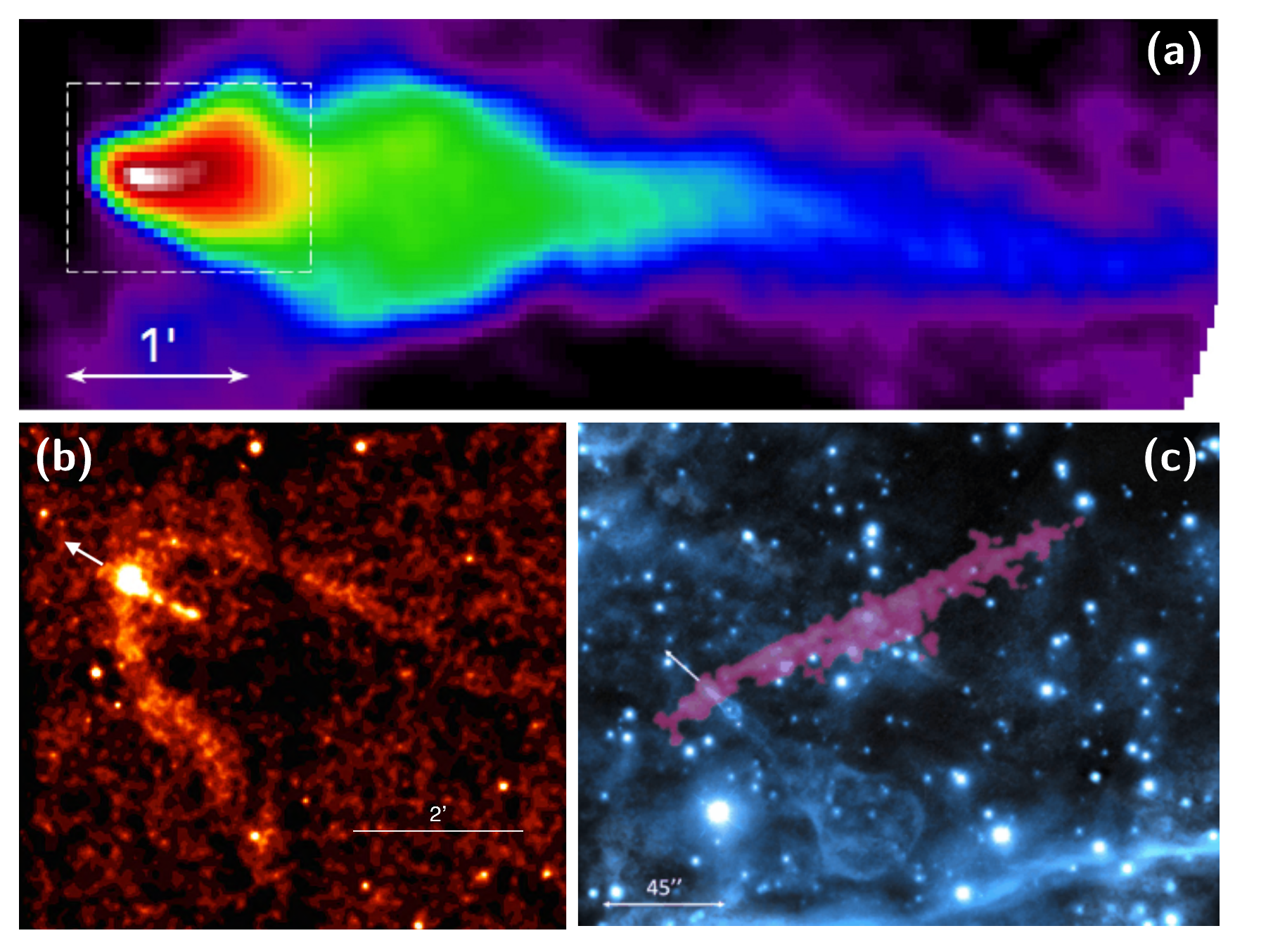}
	\caption{Selected sample of a few famous BSPWNe: (a) the Mouse nebula in radio (VLA); (b) Geminga in a combined X-ray image (Chandra-ACIS, 0.5-8 keV, 540 s); (c) the Guitar nebula and its mysterious misaligned outflow in a combined H$\alpha$ X-ray image (Chandra-ACIS, 0.5-8 keV, 195 ks) taken from the Chandra archives.}
	\label{fig:BSPWNe_OBS}
\end{figure}

In Fig.~\ref{fig:BSPWNe_OBS} we present a selection of BSPWNe to highlight the variety of the typical morphologies that are observed:
\begin{itemize}
\item The  Mouse nebula was first observed in a radio survey of the Galactic center \citep{Yusef-Zadeh:1987} and it shows one of the most extended radio tails ever seen \citep{Gaensler:2004,Hales_Gaensler+09a}. In the head it is  quasi-conical with half aperture angle of $\sim 25^\circ$ and gets narrower at a distance of $\sim 1^\prime$ behind the pulsar. It is also one of the few BSPWNe for which we have polarimetric information, suggesting a magnetic field wrapped around the bow shock head and then becoming parallel to the pulsar motion in the tail
\citep{Yusef-Zadeh:2005}. Interestingly X-rays show a more compact tail, more than a factor of ten fainter than the head, with signs of diffuse emission in a halo ahead of the pulsar itself. Deep observations in the X-ray band have been presented recently by \citet{Klingler:2018}, showing a clearer picture of the PWN. The tail shows  an evident narrowing with the distance from the pulsar and a smaller lateral expansion with respect to the radio structure.
\item The PWN associated to Geminga, on the other hand, is only observed in X-rays. It shows an asymmetric three-tail morphology with a long central tail
\citep{Posselt_Pavlov+17a}, apparently formed by isolated blobs, surrounded by two lateral tails \citep{Caraveo:2003, Pavlov:2006, Pavlov_Bhattacharyya+10a}, which show a hard spectrum with no signs of synchrotron cooling, and are not just due to limb brightening.
\item A peculiar case is that of the Guitar nebula, observed in H$\alpha$ \citep{Cordes:93, Dolch:2016}. 
Different attempts to find an X-ray counterpart compatible with the revealed H$_\alpha$ morphology have failed, while high resolution
observations made with Chandra revealed, on the contrary, a misaligned X-ray outflow, inclined by $\sim 118^\circ$ with respect to the direction of the
pulsar motion \citep{Wong:2003, Hui_Becker07a, Johnson:2010}.  
The H$_\alpha$ shape also presents a peculiar ``head-and-shoulder'' configuration, with an evident broadening with distance from
the pulsar, possibly the evidence of the mass loading of neutrals from the ambient matter \citep{Morlino:2015, Olmi:2018}. Recently a similar X-ray feature has been also seen in the Lighthouse Nebula \citep{Pavan_Bordas+14a, Marelli:2019}.
\end{itemize}

Recent observations have also revealed extended TeV halos surrounding some BSPWNe \citep{Abeysekara:2017}, though to be the signature of the
escape of high energy particles. If so,  one could use them to constrain the contribution of PWNe to leptonic antimatter in the Galaxy \citep{Blasi:2011, Amato:2018}. 

The firsts numerical models of BSPWNe dated from the past decade \citep{Bucciantini:2002, van-der-Swaluw:2003, Bucciantini:2005,
  Vigelius:2007}. 
By using multidimensional codes it was possible to extend  the simple analytical or semi-analytical models \citep{Bandiera93a,Wilkin:1996} to account for the presence of magnetic field or anisotropy in either the wind or the ISM. 
However only recently results from the first 3D simulations of BSPWNe in the fully relativistic MHD regime were presented by \citet{Barkov:2019},
where the authors investigate the morphology resulting from a few different assumptions for the magnetic field geometry and properties of the
ambient medium. 
At the same time in \citet{Olmi:2019}, Paper I hereafter, we presented a large set of 3D relativistic MHD simulations performed with adaptive mesh refinement (AMR) to improve the numerical resolution at the bow shock head, in an attempt to sample as much as possible the parameter space
characteristic of these systems.
Different  models for the pulsar wind were taken into account, implementing both isotropic and anisotropic distribution of
the energy flux, with diverse values of the initial magnetization, and defining a set of various geometries by varying the inclination of the pulsar spin-axis with respect to the pulsar kick velocity. 
In Paper I we analyzed the effects of the variation of the pulsar wind properties on the global morphology of the BSPWN, and on
its dynamics, with particular attention to the development of turbulence in the tail.

This paper is the  follow up of our previous work. Here we present emission and polarization maps computed on top of our previous simulations, focusing the discussion on the possible observational signatures. In particular we will try to assess the role of turbulence in the emission properties, the possible way to distinguish laminar versus turbulent flows, and how to use this information to guess the geometry of the system. 
Recently a simplified emission model for purely laminar flows has been presented by \citet{Bucciantini:2018}, also used to evaluate the possible escape of high energy particles in \citet{Bucciantini:2018b}, and we will compare our result with those predictions.
 
The paper is organized as follows: in Sec.~\ref{sec:methods} we recall the different models considered in our numerical analysis from Paper I, and briefly describe the methods used to compute emission and polarimetric maps; in Sec.~\ref{sec:analysis} we present the complete analysis of the results, comparing them with previous models; in Sec.~\ref{sec:conclusions} we sum up our findings and draw our conclusions.

\section{Methods}\label{sec:methods}

\begin{figure}
	\includegraphics[width=.5\textwidth]{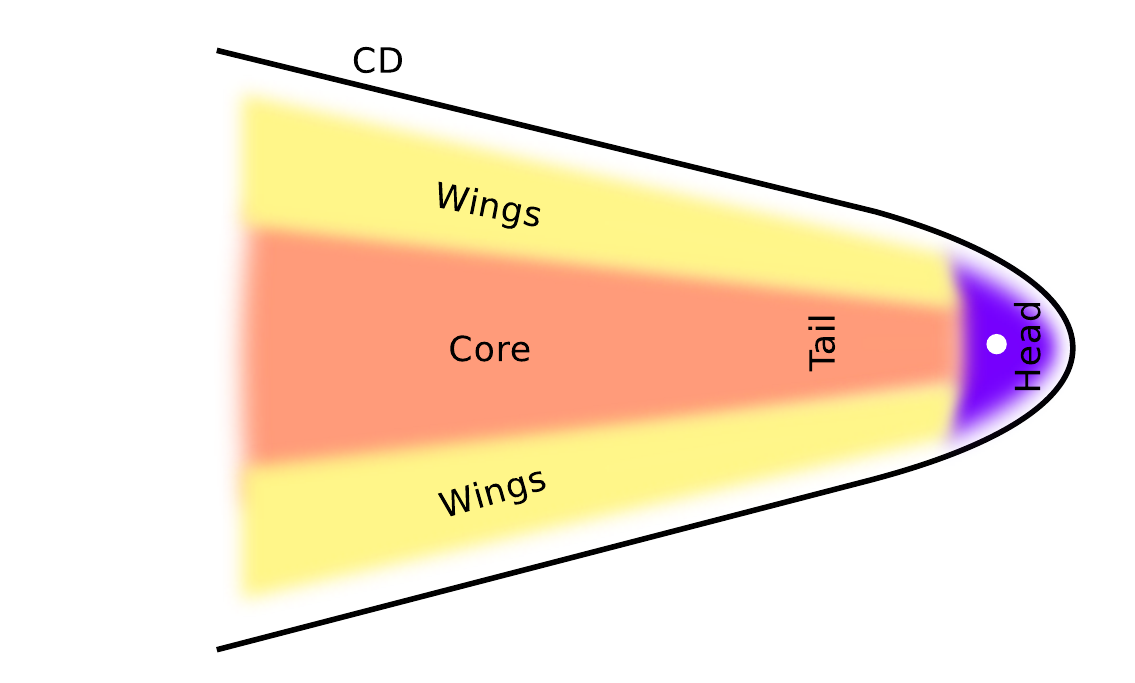}
	\caption{Scheme of the BSPWN structure: the nebula is divided into the head, the region surrounding the PSR, and the tail. The last one is further diveded in three different regions: a central core surrounded by wings, that allow an easier interpretation of the emitting and polarimetric properties.}
	\label{fig:0}
\end{figure}

\begin{figure}
	\includegraphics[width=0.4\textwidth]{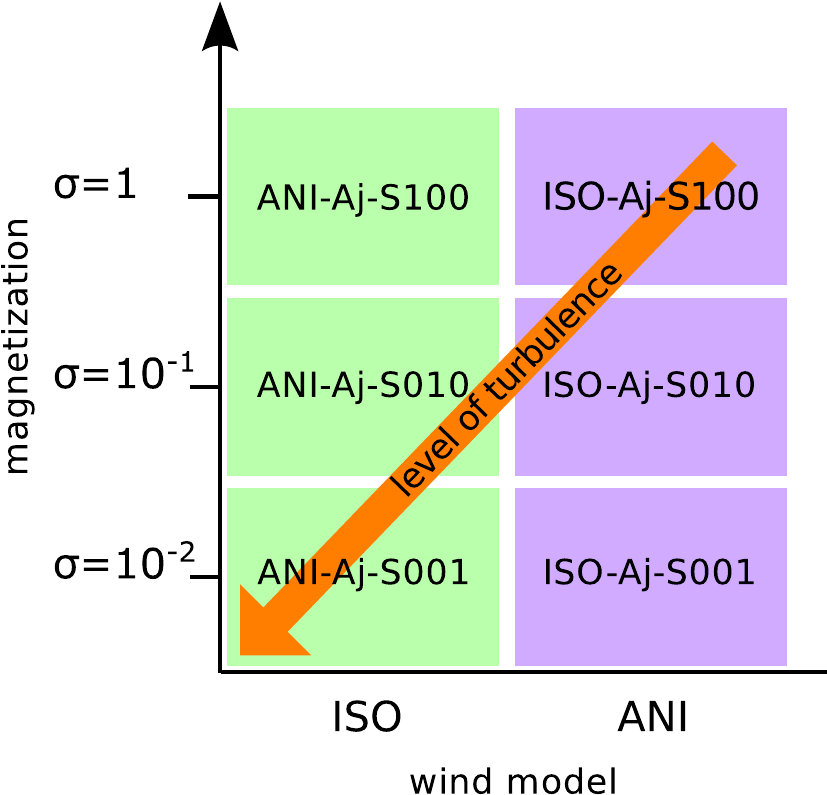}
	\caption{
	Notation used for the emission maps and its link with the value of magnetization and isotropy/anisotropy of the energy flux in the wind. ISO-A$j$-S$x$ and ANI-A$j$-S$x$ refer to the model  I$_{\{\phi_M,\, i\}}$ and  A$_{\{\phi_M,\, i\}}$ of Paper I, with $x=10^{-i}$ referring to $\sigma=[0.01,\, 0.1,\, 1.0]$ and $j = 180(\phi_M/\upi)$ to the spin to
kick inclination in degrees.
	This scheme also gives the indication of the development of turbulence in the different models (the orange arrow).
	}
	\label{fig:models}
\end{figure}

%
The 3D structure of the velocity and magnetic field is provided by our 3D Relativistic MHD (RMHD) models, developed and presented in Paper I, to
which the reader is referred for a general discussion of the setup. 
Given that in all of those models the ISM was assumed to be unmagnetized, the magnetic field, injected by the pulsar wind, can be taken as a good tracer of the relativistic material coming from the pulsar itself, and defining the part of the tail that is responsible for non-thermal emission. 
Moreover we will concentrate here on radio emission, and  radio emitting particles have synchrotron lifetime longer that the flow time in the nebula, so that we can neglect cooling. 
Let us note here that given the typical high flow speed found in numerical simulations, even X-ray emitting particles are only marginally
affected by cooling \citep{Bucciantini:2005}, such that our results can reasonably apply even to higher energies.
\begin{figure}
	\includegraphics[width=0.5\textwidth]{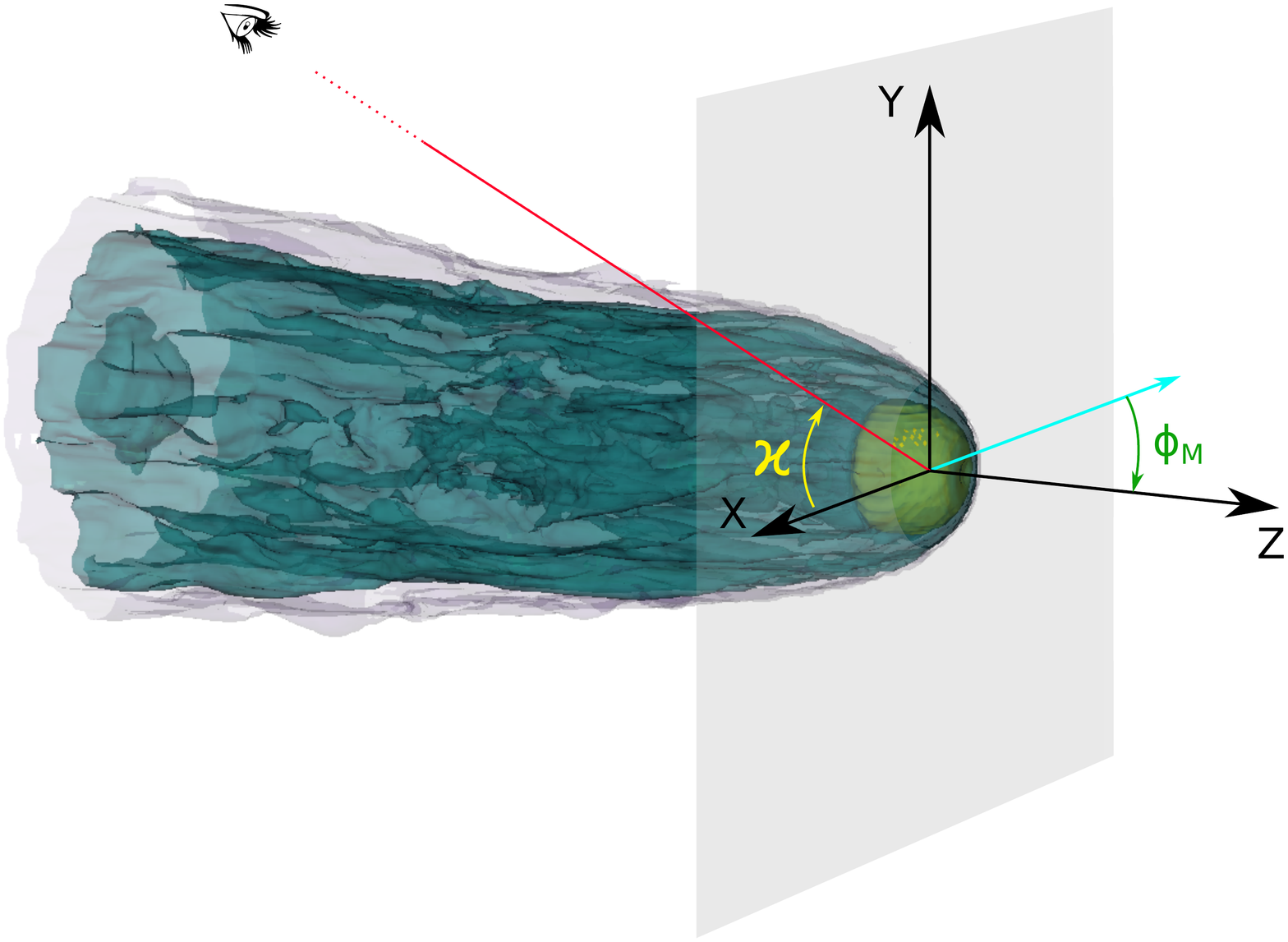}
	\caption{Representation of the angles considered in the discussion, with $\phi_M$ the inclination of the pulsar spin-axis with respect to its direction of motion (the $z$ axis) and $\mathcal{X}$ the inclination of the observer's line of sight in the $x-y$ plane, superimposed on an isolevel model of a BSPWN made with the opensource analysis tool VisIt \citep{Childs:2012}.
	}
	\label{fig:angles}
\end{figure}

 We assume, following \citet{Del-Zanna:2006}, that the emitting pairs are distributed according to a power-law in the energy $\epsilon$, $n(\epsilon)=K \epsilon^{-(2\alpha+1)}$, where $K$ is proportional to the local density of emitting particles, and their local synchrotron emissivity at frequency $\nu$, toward the observer, is
\begin{align}
	j(\nu,\boldsymbol{n})= C |\boldsymbol{B}'\times
	\boldsymbol{n}'|^{\alpha+1} D^{\alpha+2} \nu^{-\alpha}\,,
	\label{eq:spectral}
\end{align}
where $\boldsymbol{B}'$ and $\boldsymbol{n}'$ are, respectively, the magnetic field and the observer direction measured in the frame comoving with the
flow, while $C$ is a normalization constant dependent on $K$. 
$D$ is the Doppler boosting coefficient
\begin{align}
D=\frac{\sqrt{1-\beta^2}}{1-\boldsymbol{\beta}\cdot \boldsymbol{n} }=
  \frac{1}{\gamma (1-\boldsymbol{\beta}\cdot \boldsymbol{n})}\,,
\end{align}
where $\gamma$ is the Lorentz factor of the flow,  $\boldsymbol{\beta}$ and $\boldsymbol{n}$, respectively, the flow speed normalized
$c$, and the observer direction, both measured in the observer frame. Now, in terms of quantities measured in the observer frame
(unprimed), one has
\begin{align}
 |\boldsymbol{B}'\times
\boldsymbol{n}'|
=\frac{1}{\gamma}\sqrt{B^2-D^2(\boldsymbol{B}\cdot\boldsymbol{n})^2+2\gamma
D (\boldsymbol{B}\cdot\boldsymbol{n})(\boldsymbol{B}\cdot\boldsymbol{\beta})}\,.
\end{align}
One can also compute the polarization angle $\xi$, that enters into
the definition of the Stoke's parameters $Q$ and $U$ \citep{Bucciantini:2018a}. Choosing a Cartesian
reference frame with the observer placed in the $X$ direction, one finds
\begin{align}
\cos{2\xi} =\frac{q_Y^2-q_Z^2}{q_Y^2+q_Z^2},\quad\quad \sin{2\xi} =-\frac{q_Yq_Z}{q_Y^2+q_Z^2}\,,
\end{align}
where
\begin{align}
q_Y=(1-\beta_X)B_Y+\beta_Y B_X\,,\quad\quad q_Z=(1-\beta_X)B_Z+\beta_Z B_X\,.
\end{align}

\begin{figure*}
        \includegraphics[width=1.\textwidth]{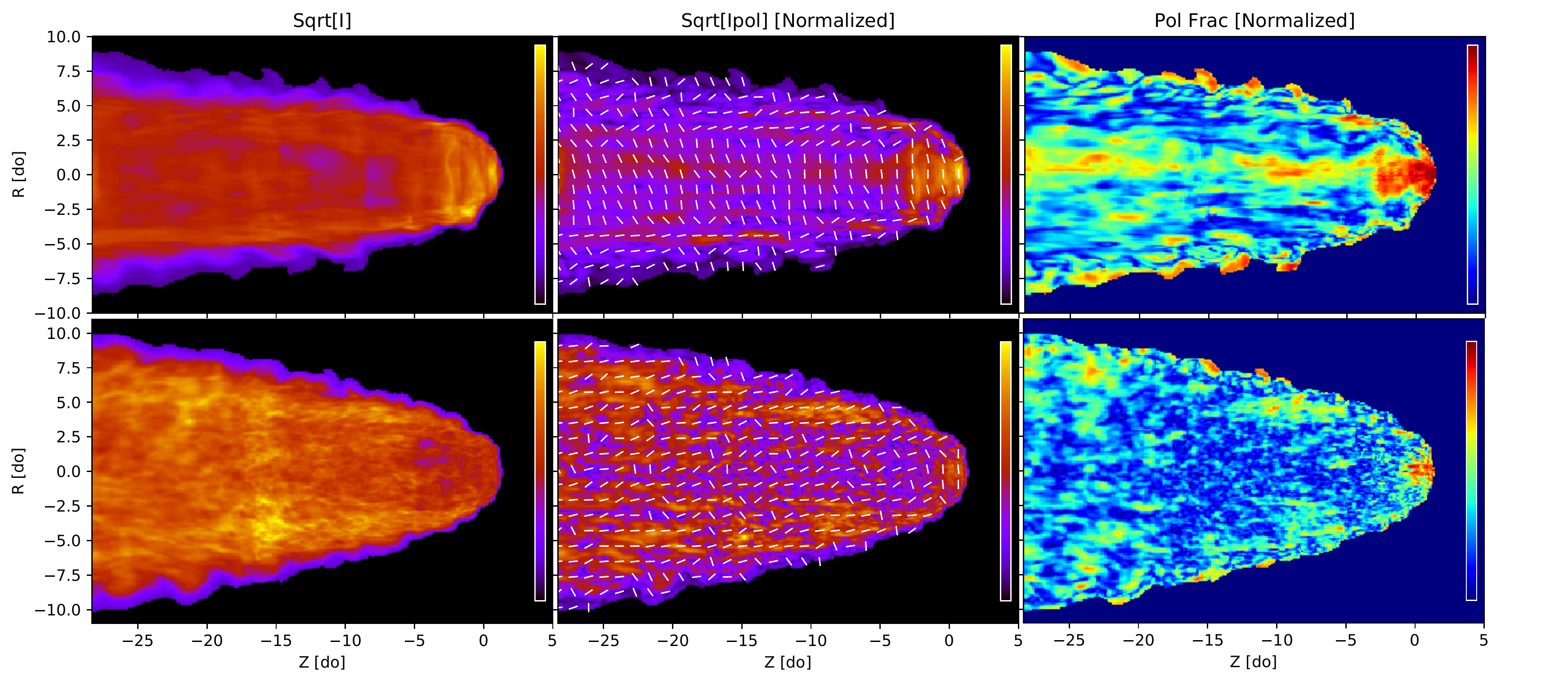}
	\caption{Maps for the case UNI-A00-S100 (upper row) and case UNI-A00-S001 (bottom row), with uniform local emissivity. From left to right: square root of the total synchrotron intensity normalized to the maximum, square root of the  polarized intensity normalized to the maximum superimposed with the polarized direction, and polarized fraction PF, normalized to the theoretical maximum for a power law synchrotron with $\alpha=0$. The color scale is linear between zero and one.  }
	\label{fig:1}
\end{figure*}

In general the local density of emitting particles $K$, as well as the power-law index $\alpha$, might differ in different locations, depending on how particles are injected. 
For simplicity we assume that the power-law index $\alpha$ is uniform in the nebula. For the emitting particle density we adopt
two different choices: either a uniform distribution, as was done in \citet{Bucciantini:2018}, or a density proportional to the local values of the thermal pressure,  as it is customary in other PWNe models \citep{Volpi:2008, Olmi:2014}. We have also investigated a third possibility that the emission is concentrated in the current sheets that form in
the BSPWN. In this case their local synchrotron emissivity at
frequency $\nu$, toward the observer, is taken to be
\begin{align}
	j(\nu,\boldsymbol{n})= C_{\rm J} \|\boldsymbol{J}\| D^{\alpha+2} \nu^{-\alpha}\,,
\label{eq:spectral}
\end{align}
where $\boldsymbol{J}$ is the  current density, computed as the curl
of $\boldsymbol{B}$ (we have verified that the displacement current is
negligible), and $C_{\rm J} $ is just a normalization constant which
we assume to be position independent. In this case we also compute polarization just by
taking as the polarization direction the one given by the local magnetic field. Note that ohmic
dissipation should scale as $\|\boldsymbol{J}\|^2$, however because of
the use of AMR, given that the resistivity is purely numerical in
nature, such choice leads to artificial jumps in the surface
brightness in coincidence with AMR refinement boundaries, given that $\|\boldsymbol{J}\|^2$
it is not volume conserved. This spurious
effect is strongly mitigated if one takes the simple norm of the
current, given that this quantity is instead volume conserved in ideal MHD.

\section{Emission maps}\label{sec:analysis}

We build all of our maps using a spectral index $\alpha=0$. \citet{Bucciantini:2018}
showed that this is a good average for the observed radio spectra, and
changes in the typical observed range do not affect much the results.
The polarized fraction is alwyas given in terms of the theoretical
maximum, that  is $0.6$ for $\alpha=0$. To keep the discussion as
simple as possible, we have schematized the BSPWN as shown in
Fig.~\ref{fig:0}, highligthening the key regions whose properties we are
going to discuss, and compare.

Emission maps where computed for all simulations presented in Paper I. We adopt here the same notation for the names of the
different reference RMHD models, while for the emission maps we use a different notation, which allows one to
identify them  in terms of the physical parameters. 
The new notations ISO-A$j$-S$x$ and ANI-A$j$-S$x$ refer to the model  I$_{\{\phi_M,\, i\}}$
and  A$_{\{\phi_M,\, i\}}$ of Paper I, with $x=10^{-i}$ (referring to $\sigma=[0.01,\, 0.1,\, 1.0]$) and $j = 180\phi_M/\upi$ (the spin-to-kick inclination in degrees) as shown Fig.~\ref{fig:models}, where we show also the trend toward more turbulent configurations.

In Fig.~\ref{fig:angles} we illustrate the typical viewing geometry we use to build our maps. We sampled the full range of possible inclination angles $\mathcal{X} =[0,360]^\circ$. However as it was done in \citet{Bucciantini:2018}, to keep the discussion as simple as possible, focusing only on the major trends, we limited our presentation here to few values of $\mathcal{X} $, typically $[0,45,90]^\circ$, being the ones for which we do expect the largest
differences. Moreover we consider only a viewing geometry where the PSR kick velocity lays on the plane of the sky.

\subsection{Uniform wind luminosity}

We begin our analysis with the fully axisymmetric case $I_{\{0,\,0\}}$, corresponding to an isotropic energy injection and a high magnetization value  $\sigma=1$. The dynamics in this case shows a low level of turbulence, that only develops far away from the pulsar.  For the sake of comparison with the laminar models, we first discuss the case of a uniform emissivity, deferring the case of an emissivity scaled by the pressure to the end, because the different choices in the normalization of the local emissivity mostly affect the emission maps, and not as much the polarization properties.

In this case the luminosity peaks in the head region, and is almost constant without any evidence for limb brightening (Fig.~\ref{fig:1}). 
The wings of the tail are about a factor of two brighter with respect to the core. The total emission has in this case a {\it jellyfish} like morphology.
These results are in reasonable agreement with the laminar models, except that there is no evidence for the slower region downstream of
the Mach disk, that was found in the latter ones.

The polarized fraction is higher in the head, whose central part reaches values $\sim 0.9-0.8$, with edges down to $\sim 0.2-0.3$. There is some indication that the polarization rises again very close to the contact discontinuity (CD) up to values  $PF\sim
0.7-0.9$, but this might be due to spurious effect associated with the low surface brightness of those regions.
In the tail the core has an average polarized fraction of $0.5$ with local peaks up to $0.8$, while the wings are depolarized ($ PF\sim 0.2-0.3$).
This trend is in accordance with the findings of laminar models, even if the development of turbulence strongly reduces the
polarized fraction. The polarized angle in the head is similar to the one of the laminar models, the tail has a core with the polarization orthogonal to the nebular axis, as expected for a toroidal field injected by a pulsar with spin-axis aligned with the kick velocity, while in the wings it tends to become aligned with the tail. This is due to a combination of local shear at the CD and relativistic polarization angle swing, that was already noted in the laminar case.

As the magnetization becomes smaller, Fig.~\ref{fig:1} bottom row, and turbulence increases,
there is a change in the emission properties. For the case $I_{\{0,\,2\}}$, corresponding to a wind magnetization $\sigma=0.01$, the luminosity is now
more uniform with no apparent distinction between the core and the wings; the head tends to be under-luminous; very fine structures appear in the luminosity maps on scales $\ll d_0$, probably related to the small scale turbulence that is seen in the velocity maps (see Paper I). 
Here $d_0=[\dot{E}/(4\upi c \rho_\mathrm{ISM} v_\mathrm{PSR}^2)]^{1/2}$ is the stand-off distance, the typical length scale of a BSPWe, depending on the pulsar luminosity ($\dot{E}$), kick-velocity ($v_\mathrm{PSR}$) and density of the ambient medium ($\rho_\mathrm{ISM}$), with $c$ the speed of light.
The nebula is in general unpolarized, with perhaps some marginal residual polarization in the very head. The polarization maps again show the fine structures related to turbulence observed in the emission maps. There is a slightly higher polarized flux in the wings toward the CD. In general, no appreciable variation is seen with the inclination angle of the observer.

On the other hand if one assumes that the local emissivity  scales with the local value of the gas pressure, the emission maps change substantially, while the results concerning the polarization properties do not as much, suggesting that it is just the structure of the magnetic field, more than the particle energy
distribution, that regulates the polarization properties. 
 \begin{figure}
	\includegraphics[width=.5\textwidth]{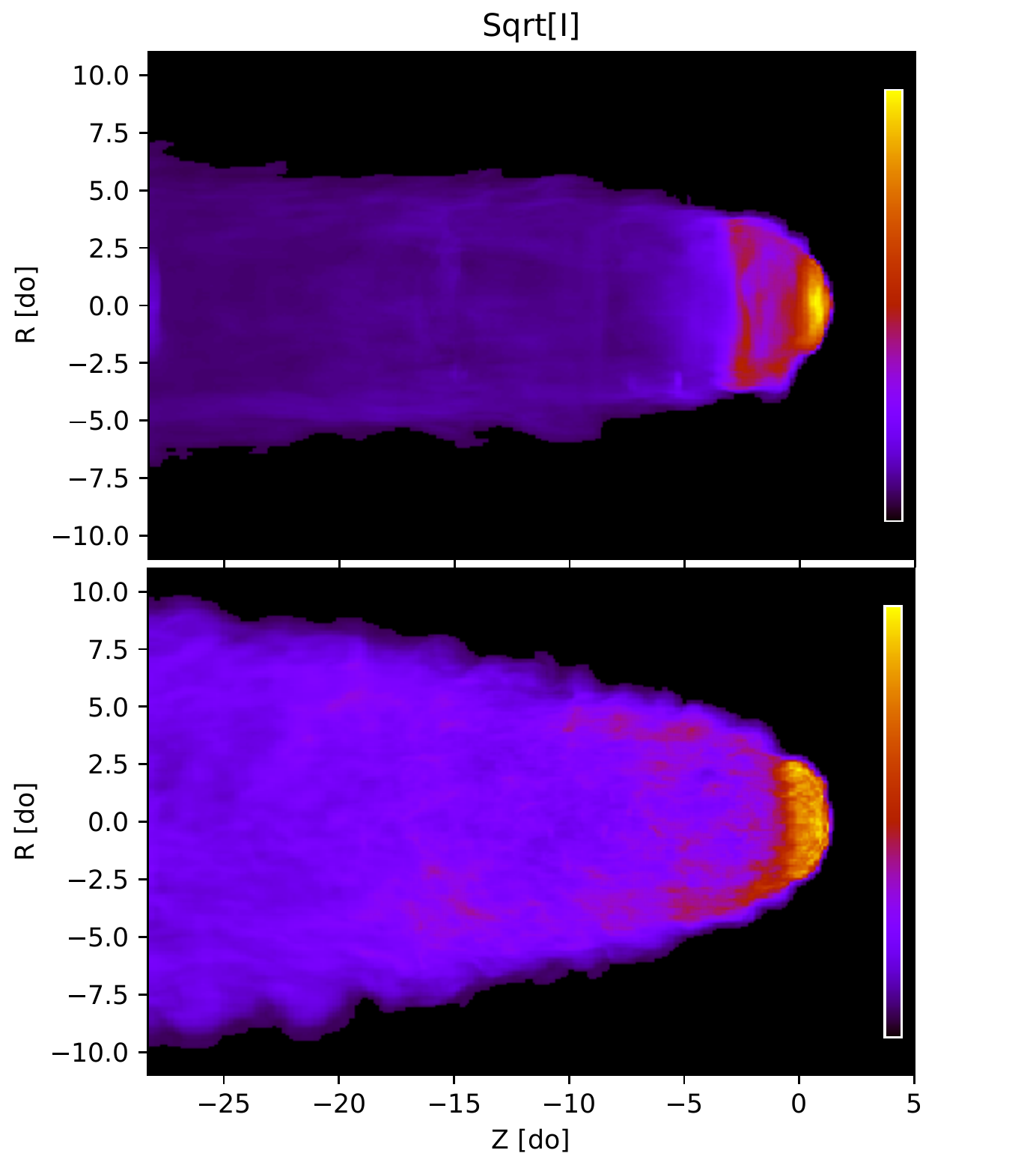}
	\caption{Maps computed assuming the local emissivity scales with the local value of the gas pressure. 
         Square root of the total synchrotron intensity: upper panel for the  UNI-A00-S100 case (to be compared with Fig.~\ref{fig:1} upper panel), bottom panel
         for the  UNI-A00-S001 case (to be compared with Fig.~\ref{fig:1}, bottom panel), both normalized to their maximum.}
	\label{fig:3}
\end{figure}

\begin{figure*}
	\includegraphics[width=1.0\textwidth,clip]{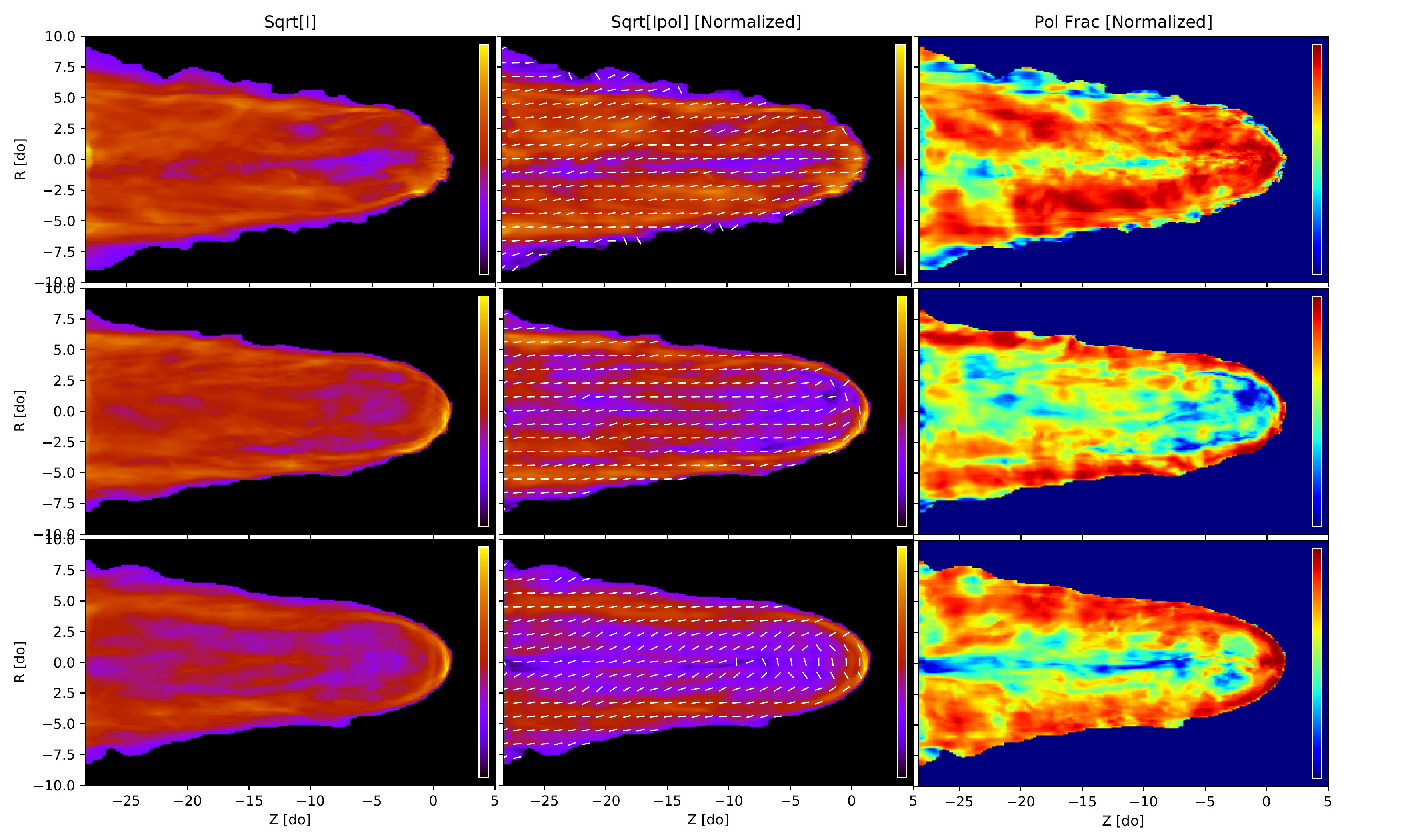}
	\caption{Same as Fig.~\ref{fig:1} but here  for the case UNI-A90-S100: top row for a viewing angle $\mathcal{X} =0^\circ$,
          middle row $\mathcal{X}=45^\circ$, bottom row $\mathcal{X}=90^\circ$. }
	\label{fig:4}
\end{figure*}
In Fig.~\ref{fig:3} we show the total intensity for the cases $I_{\{0,\,0\}}$ and case $I_{\{0,\,2\}}$, with the emissivity now weighted by the local pressure.
The total emission, as well as the polarized emission,  if far more enhanced in the head, where the pressure can be a factor ten or
more higher than in the tail. In this case the luminosity is dominated by the very front part of the head, which can easily be a factor 10-20 brighter than the
tail, independently on the magnetization. The same holds in the low $\sigma$ case, where it is the very front of the bow shock that dominates
the emission. The tail however is now slightly brighter in comparison to the high $\sigma$ case, partly reflecting the trend seen for a uniform emissivity, with no clear distinction between core and wings.

In the perpendicular case  $\phi_M=\upi/2$, instead we have analyzed the emission properties for various values of the observer inclination angle $\mathcal{X}$. We begin again by discussing the case corresponding to a high magnetization $\sigma =1$  and a uniform emissivity (case $I_{\{\upi/2,\,0\}}$),  
which has a more laminar structure and can be compared with those in \citet{Bucciantini:2018}. A set of maps computed for selected values of $\mathcal{X}$  is shown in Fig.~\ref{fig:4}.

\begin{figure*}
	\includegraphics[width=1.0\textwidth,clip]{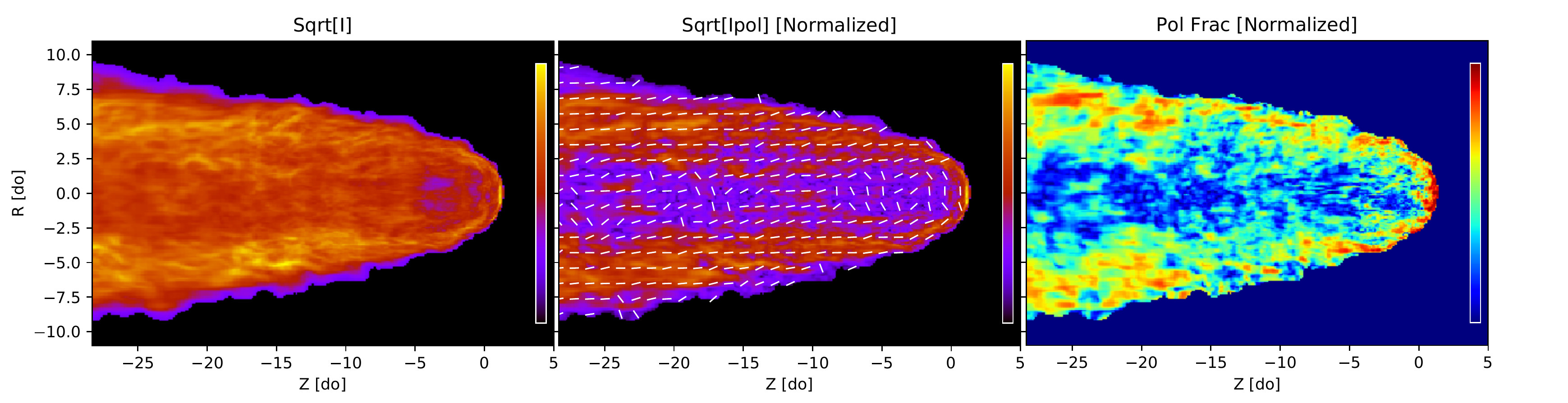}
	\caption{Same as Fig.~\ref{fig:1} but here for the case  UNI-A90-S001, with $\mathcal{X}=90^\circ$.  }
	\label{fig:5}
\end{figure*}
We find that for all the possible values of $\mathcal{X}$, the tail brightness is higher in the wings than in the core by about a factor 2. 
The difference is more pronounced at $\mathcal{X}=90^\circ$. The head instead shows larger changes with the observer inclination. At
$\mathcal{X}=0^\circ$ it approximatively has the same intensity of the tail, and looks quite uniform. As $\mathcal{X}$ increases to $90^\circ$ it becomes brighter, with an enhanced limb right at the location of the CD, marking the position of the shocked pulsar wind in the head, between the TS and the CD itself. In terms of polarization properties the results are now reversed with respect to the aligned case: the polarized fraction is much higher in the wings and in the head  with values up to $0.9$, and reduced in the core of the tail which is almost depolarized for $\mathcal{X}=90^\circ$. For $\mathcal{X}=45^\circ$ the head looks slightly depolarized.
The polarized angle instead looks fully aligned with the tail, with deviation of at most  $\pm 15^\circ$, except in the $\mathcal{X}=90^\circ$ case where in the head it looks orthogonal. These results are again consistent with those found using laminar models.

Lowering the magnetization, the flow becomes more turbulent and the overall appearance of the nebula, both in terms of total intensity and
polarized properties, changes. 
For $\sigma =0.01$, case $I_{\{\upi/2,\,2\}}$ (see Fig.~\ref{fig:5}), the nebula looks quite similar to the aligned case: the bulk of the head is less luminous than the tail, there are small scales structure, and the level of polarization is small. 

There is some residual evidence that the wings are slightly brighter than the core, and slightly more polarized (this is more evident for $\mathcal{X}=90^\circ$). As expected for highly developed turbulence, the appearance and polarization properties do not depend much on the inclination angle with respect to the observer. At $\sigma=0.1$, case $I_{\{\upi/2,\,1\}}$, we have an intermediate regime. The head has the same brightness as the core of the tail, while the wings are slightly brighter.  
In general for lower magnetizations we observe smaller scales in the polarization maps, and the polarized fraction has local patches as high as $PF=0.8$, with an average of about $0.6$.

\begin{figure}
	\includegraphics[width=.5\textwidth,clip]{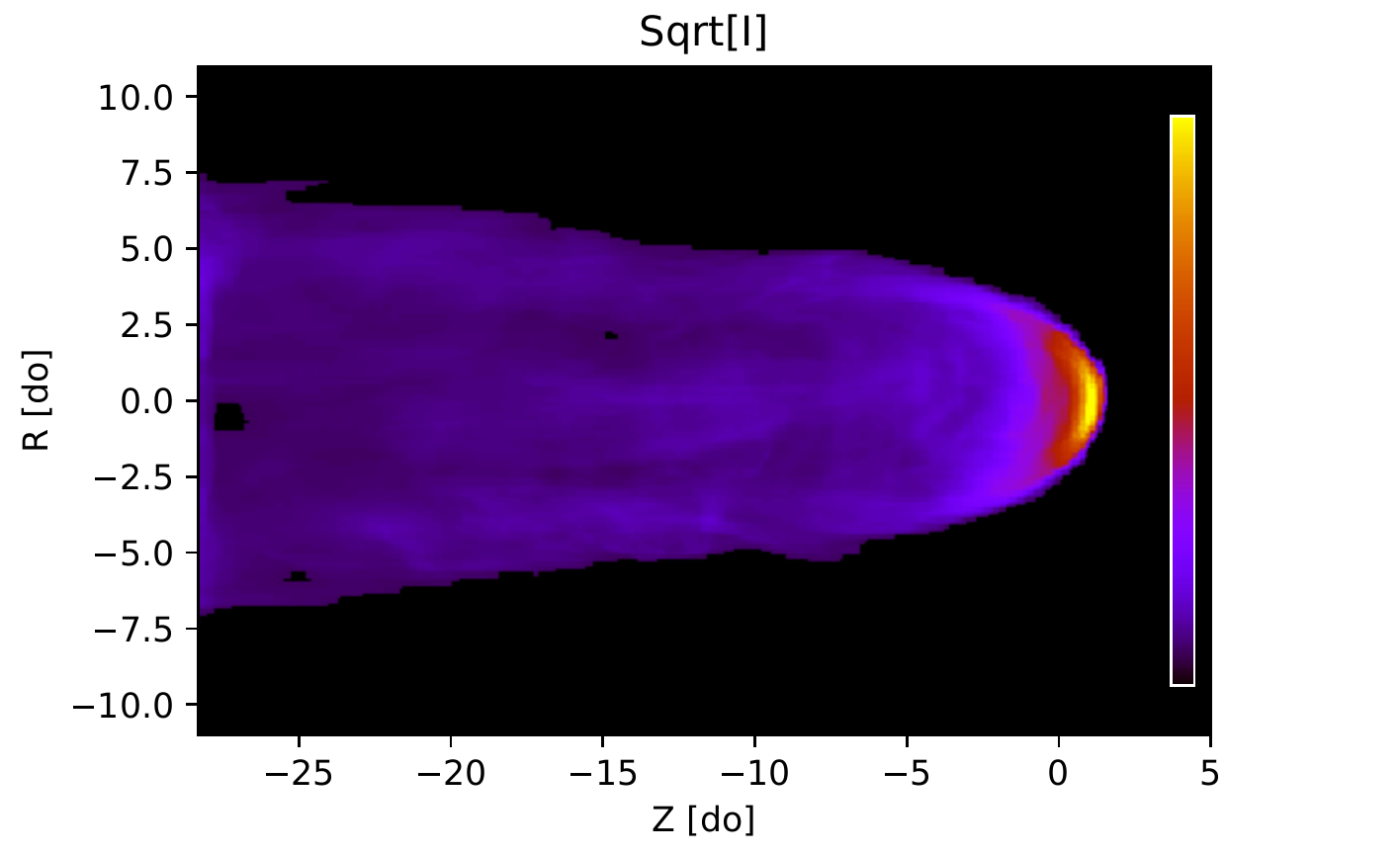}
	\caption{Map of the square root of the intensity computed assuming the local emissivity scales
          with the local value of the gas pressure, for the UNI-A90-S100 case, with a viewing angle $\mathcal{X}=90^\circ$ 
          }
	\label{fig:6}
\end{figure}

If instead one  considers cases with the emissivity scaling with the pressure, one finds again that the polarized properties are essentially not much affected.
What changes mostly is the emission pattern that now is dominated by a bright region in the very head, corresponding to the shocked layer between the termination shock and the CD, as shown in Fig.~\ref{fig:6} for the case with the highest magnetization and $\mathcal{X}=90^\circ$. 

This high brightness region is more pronounced in the high $\sigma$ regime than at lower magnetizations, where the head-to-tail brightness ratio tends to
be smaller.

\begin{figure*}
	\includegraphics[width=1.0\textwidth,clip]{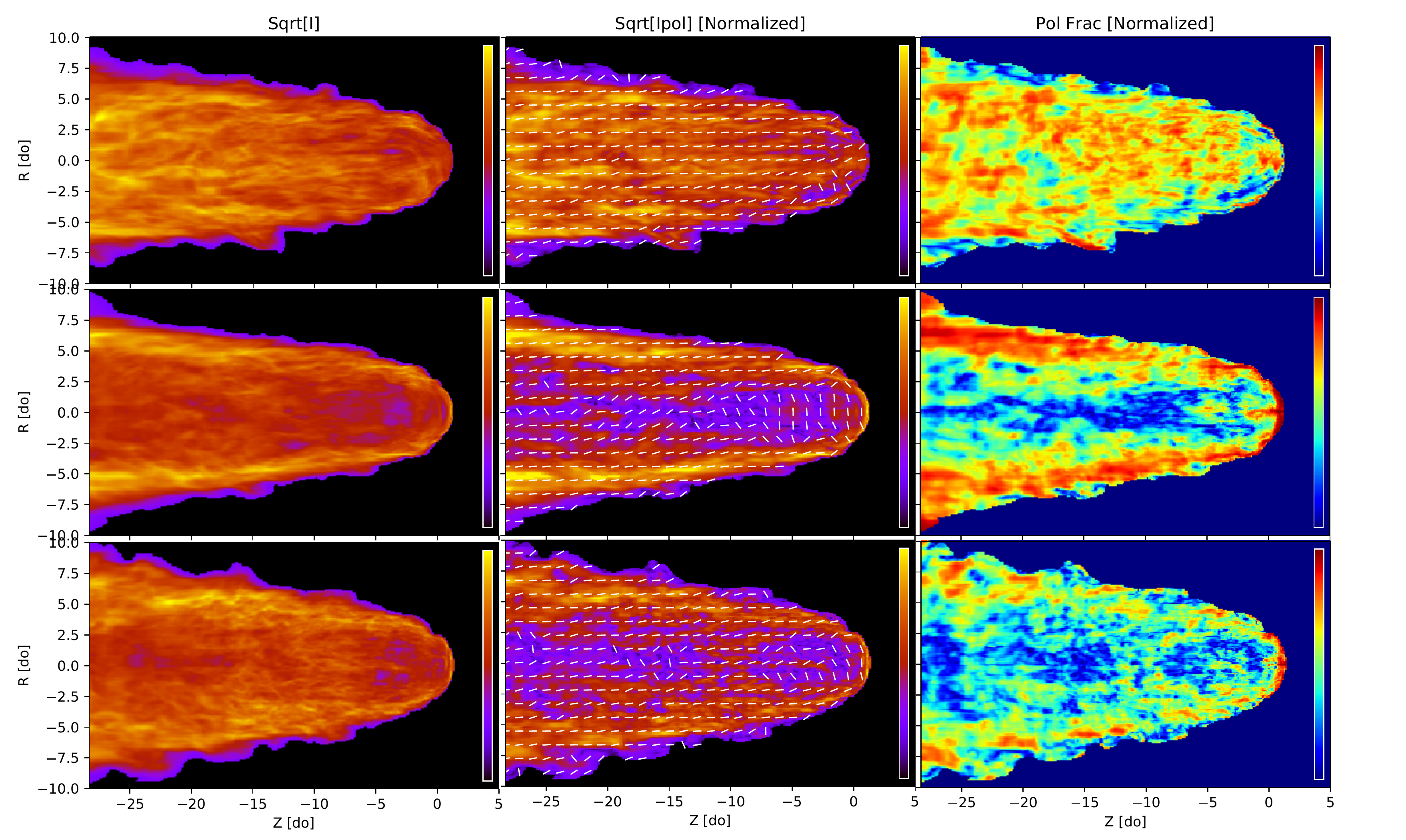}
	\caption{Same as Fig.~\ref{fig:1}: top row case UNI-A45-S100 for a viewing angle $\mathcal{X} =0^\circ$, middle row case UNI-A45-S100 for $\mathcal{X}=90^\circ$, bottom row case UNI-A45-S001 for $\mathcal{X}=90^\circ$. }
	\label{fig:7}
\end{figure*}

The case of a pulsar spin-axis inclined by $\phi_M=45^\circ$ with respect to the kick velocity shows, in the high magnetization case $\sigma=1$ ($I_{\{\upi/4,\,0\}}$) a variety of patterns in the emission and polarization properties that depends on the inclination angle of the
observer (see Fig.~\ref{fig:7}). 

For $\mathcal{X}=0^\circ$ (upper row of Fig.~\ref{fig:7}) the tail appears about a factor of 2 brighter than the head, and there is no wings-to-core difference.
The polarization angle is aligned with the tail, and the polarized fraction is on average $\sim 0.6$. 
As $\mathcal{X}$ increases towards $90^\circ$, as shown in the middle row of Fig.~\ref{fig:7}, the emission pattern increasingly resembles the orthogonal
case, with brighter wings and a depolarized core. 
This is more markedly visible in the polarized intensity, which shows a very intense enhancement in the wings region and a limb brightening of the head.
Given that  the dynamics of this case is more turbulent than the previous ones, we observe the presence of lots of small structures in the maps of  the polarized fraction.
\begin{figure}
        \includegraphics[width=.5\textwidth,clip]{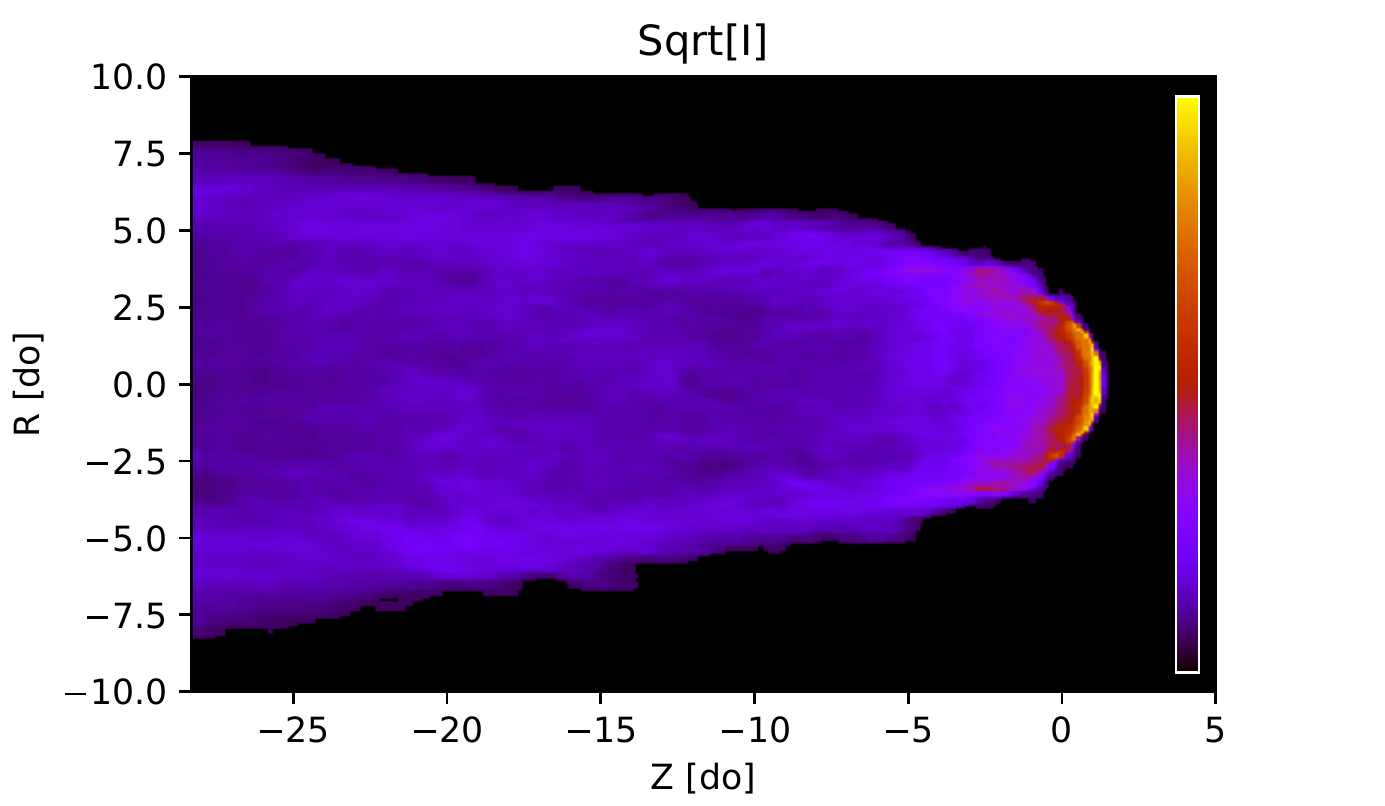}
	\caption{Map computed assuming the local emissivity scales
          with the local value of the gas pressure. 
         Square root of the total synchrotron intensity
         for the UNI-A45-S100 case and $\mathcal{X}=90^\circ$.}
	\label{fig:8}
\end{figure}
 At low magnetization $\sigma=0.01$ (bottom row of Fig.~\ref{fig:7}) the effect of turbulence is more
pronounced, and the wind-to-core difference seen at $\mathcal{X} = 90^\circ$ is far less evident. 
Polarization is in general smaller. 
Again, weighting the emissivity with the pressure leads to emission maps that are dominated by the very head, with a strong limb in the front of
the system. Interestingly when $\mathcal{X} = 90^\circ$ (see Fig.\ref{fig:8}), the orthogonal case and the inclined one show very similar emission and polarization maps.

\subsection{Anisotropic wind luminosity}
In Paper I we investigated also cases with a pulsar wind energy flux dependent on the colatitude $\theta$, according to a $F(\theta)\propto \sin^2 \theta$
dependence, as expected for force free split monopoles \citep{Michel:1973, Bogovalov:2001, Spitkovsky:2006}.
It was found that the flow in this case tends to show a higher level of turbulence,
even at high magnetization ($\sigma=1$).

\begin{figure*}
       	\includegraphics[width=1.0\textwidth,clip]{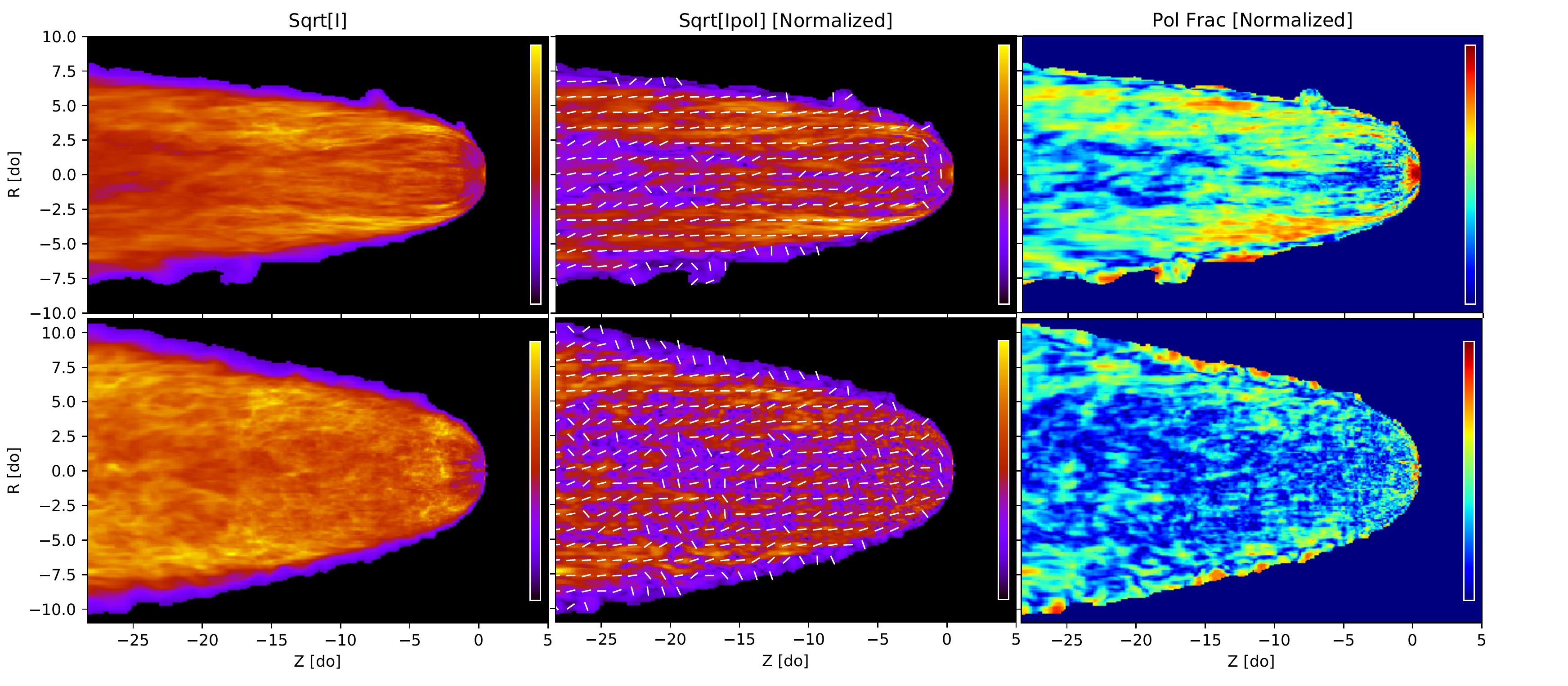}
	\caption{From left to right: square root of the total synchrotron intensity, square root of the polarized intensity, polarized fraction, all normalized in terms of their maxima. The upper row refers to case ANI-A00-S100, while the bottom one to case ANI-A00-S001. The observer's viewing angle is in both cases $\mathcal{X}=0^\circ$. }
	\label{fig:9}
\end{figure*}

We begin again discussing the high magnetization aligned case $A_{\{0,\,0 \}}$,  and the low magnetization one
$A_{\{0,\,2 \}}$, both shown in Fig.~\ref{fig:9}.
Now at $\sigma=1$ the intensity pattern looks quite different.  The head is sub-luminous with respect to the tail, and the wings are brighter than the core. 
The intensity map shows the presence of long filamentary structures, while the overall polarization is small ($PF \sim 0.3$) and there is no
evidence for either a polarized head or a polarized core, as it was found for the isotropic wind luminosity. At low magnetization ($\sigma
=0.01$) instead the emission looks very similar to the isotropic case: the tail is brighter than the head; the wings are not markedly distinct from the core; the nebula is almost completely depolarized. 
This same considerations apply also if one considers a local emissivity that scales as the local gas pressure. 
The only evident variation with respect to the isotropic injection case is that now the emission in the head is weaker, and the head-to-tail difference is
less pronounced.

\begin{figure*}
	\includegraphics[width=1.0\textwidth,clip]{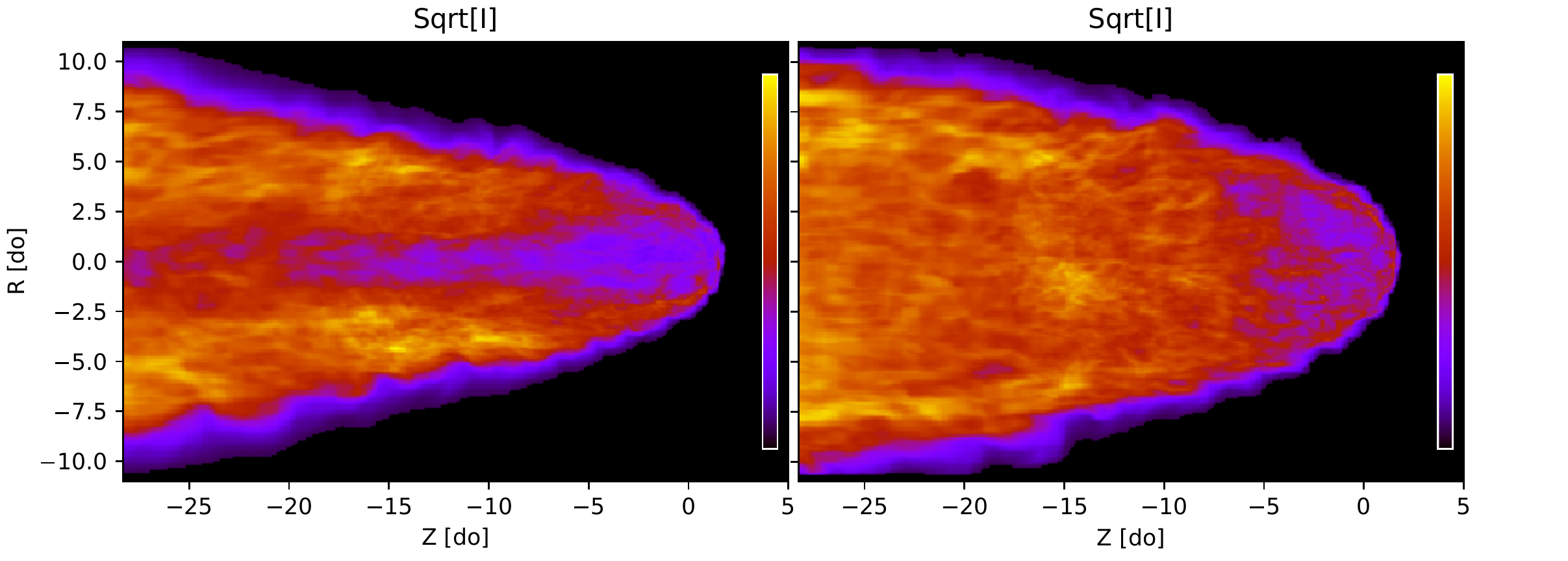}
	\caption{Square root of the total synchrotron intensity for case ANI-A90-S001, with $\mathcal{X}=0^\circ$ (left panel) and $\mathcal{X}=90^\circ$ (panel on the right). }
	\label{fig:10}
\end{figure*}

This same trends are also found for a spin-axis orthogonal to the kick velocity, $\phi_M=\upi/2$. 
The major differences with respect to the uniform injection case are that the head is basically fainter than the tail
for all the possible inclinations of the observer, and for $\mathcal{X}=0$ the wings-to-core brightness ratio tends to be stronger, as shown in
Fig.~\ref{fig:10}.

The polarization pattern is quite similar to the the uniform injection case,  but the polarized fraction is in general a factor 2 to 3 smaller, and characterized by the presence of fluctuations on small scales.  These differences holds also in the $\sigma=0.01$ case, where
now the head is markedly fainter than the tail. 
For $\mathcal{X}=0$ the wings are still brighter than the core, while for all the other observer inclination angles the tail luminosity looks quite uniform. The fact
that for an anisotropic injection the head tends to be fainter implies that when one considers a case of local emissivity proportional to the local pressure, the observed enhancement of the head luminosity with respect to the tail one is less pronounced.
\begin{figure*}
	\includegraphics[width=1.0\textwidth,clip]{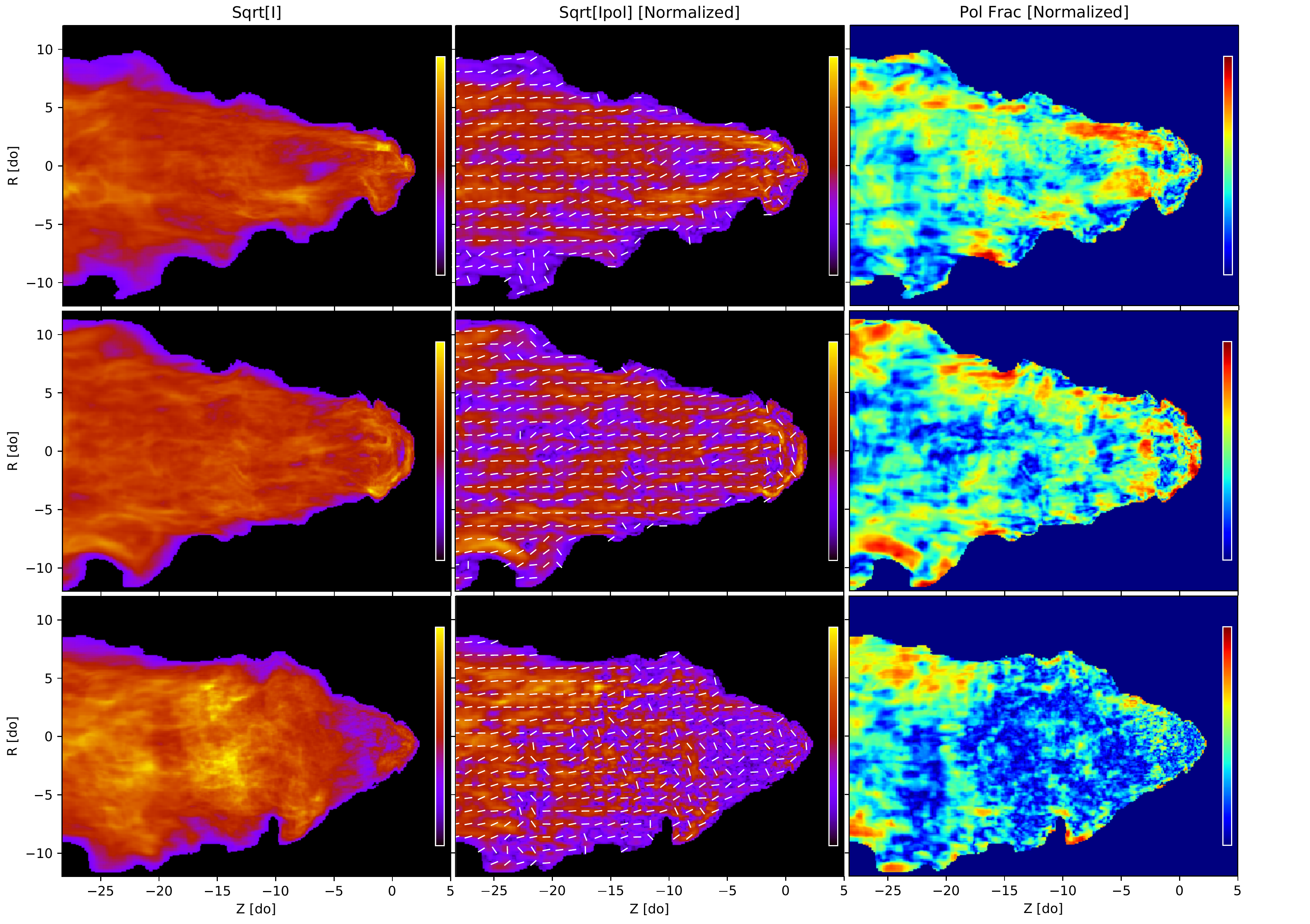}
	\caption{From left to right: square root of the total synchrotron intensity, square root of the polarized intensity, polarized fraction, all normalized in terms of their maxima. The upper row refers to the  case ANI-A45-S100  with $\mathcal{X}=0^\circ$, the middle one to  the  case ANI-A45-S100 with $\mathcal{X}=45^\circ$, while the bottom one to the case ANI-A45-S001  with $\mathcal{X}=0^\circ$. }
	\label{fig:11}
\end{figure*}

Far more interesting is the case of a spin-axis inclined by $\phi_M=45^\circ$  with respect to the pulsar proper motion, cases $A_{\{\upi/4,\, i \}}$, shown for different magnetizations and viewing angles in Fig.~\ref{fig:11}. 
It was already noted in Paper I that this configuration was by far the most turbulent one, leading to major fluctuations in the shape of the bow shock, and
triggering the formation of large blobs and waves propagating along the CD. 
All of this becomes particularly evident in the emission maps. 
The level of polarization is in general small, some residual global polarization $\sim 0.3$ is still present at $\sigma=1$, even if the
polarization pattern can be very patchy, but at lower magnetization the nebula looks almost completely depolarized. 
At $\sigma=1$ there is no difference in the intensity between the head and the tail, and overall the intensity maps do not show any distinction between wings
and core. 
At $\sigma=0.01$ the head is sub-luminous with respect to the tail even by a factor of a few. The global morphology of emission is also
quite different when the direction of observation is varied.
\begin{figure*}
	\includegraphics[width=1.0\textwidth,clip]{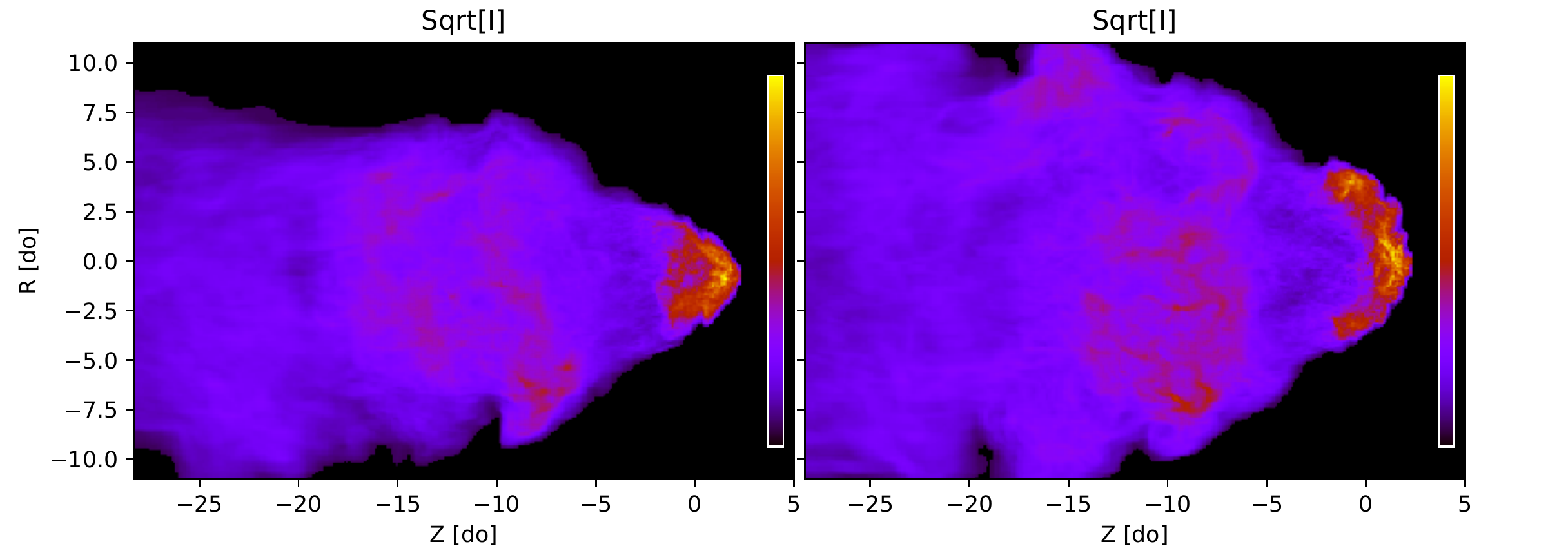}
	\caption{Square root of the total synchrotron intensity for
          cases ANI-A45-S001 $\mathcal{X}=0^\circ$ (left) and $\mathcal{X}=90^\circ$ (right). }
	\label{fig:12}
\end{figure*}

If the local emissivity is weighed by the pressure, the head becomes again the dominant feature, 
but for $\sigma=0.01$, patches in the tail with comparable luminosity can appear, as shown in Fig.~\ref{fig:12}

\subsection{Emission scaled with the currents}
It has been suggested by several authors that current sheets and
current layers could be important acceleration sites \citep{Uzdensky_Cerutti+11a, Cerutti:2014, Lyutikov:2016}, and
that they might play an important role in the origin of non-thermal
particles in the turbulent environment of young PWNe. 
Here we investigate how the emission properties change if one assumes that the local syncrotron emissivity scales with the intensity of the
local currents. 
In general the polarization properties (polarized fraction and angle) are not much affected, given that they depend of the overall magnetic
structure of the system. 
We find that there is a tendency toward the formation of fine scales in the polarized fraction pattern, and that on average the polarized fraction of the various components (wings/core/head) tends to be slightly smaller, but this effects are only appreciable in high resolution maps, while they disappear at lower 
resolutions. 
Major changes are seen mainly in the total emissivity: the head region tends in all cases to be brighter, and the emission
morphology looks to be dominated by bright limbs close to the CD for the high magnetization $\sigma=1$, and by a more uniform emissivity for the low magnetization $\sigma=0.01$. 
Interestingly, apart for the aligned case, where this limb-brightening is less evident, all the other cases show very similar emission morphology independently of the inclination angle of the pulsar spin-axis, or the orientation of the viewing angle. 
As anticipated, some effects of the AMR boxing are visible in the emission maps as changes in the average brightness, which drops as the resolution
decreases.  
This is more likely due to the fact that our code does not include explicit resistivity, and the size, thickness, and intensity
of the currents are solely regulated by numerical dissipation. 
Interestingly this effect is more pronounced in the very turbulent cases, and much less in the laminar ones, suggesting that as long as the flow remains laminar, the role of numerics in the dissipation of the magnetic field is weaker.

\section{Summary and Conclusions}
\label{sec:conclusions}
In this manuscript we have studied the emission and polarimetric
properties of BSPWNe based on the high resolution numerical models presented and
discussed in our previous work Paper I \citep{Olmi:2019},
investigating and analyzing a large set of different geometries of the BSPWNe. 

While in that work the discussion was mostly focused on the flow pattern and the development of 
turbulence in the BSPWN tail, here we investigated those
differences and their relation to the injection conditions in the pulsar wind, to see if they 
could be recovered from the properties of the emission, or if the
emissivity is or not a good tracer of the wind properties.

We focused our discussion of the possible differences among various
cases mostly to the case of high magnetization, where the flow is
quasi-laminar and the structure in the tail preserves information from
the injection region in the head, while in low magnetized
cases turbulence tends to wash out this information.

We found that in the case of uniform injection and high magnetization,
there is indeed a strong correlation between the injection properties,
as the inclination $\phi_M$, and the surface brightness of our
maps. 
The difference agrees with the expectation of fully laminar
semi-analytic models of \citet{Bucciantini:2018}. We do observe a large
variety of morphologies, from head dominated to tail dominated, with or without
bright wings, and with different structures in the polarized
fraction. 
The dependence on the viewing angle $\mathcal{X}$ is less marked
and probably only appreciable at high resolution. 
In the high $\sigma$  regime the polarized fraction can be on average quite high in the
tail.  
Once the magnetization drops we do see a drastic change towards 
a more turbulent regime. In this case it is far more difficult to
find clear observational patterns that could be used to distinguish cases
with different inclination $\phi_M$ in a robust way. 

An important characteristic of our maps is that, apart from the strongly
turbulent cases $A_{\{\upi/4,i\}}$, we do not observe major fluctuation
along the tail direction. This suggests that time variability
associated to temporal changes in the flow pattern should not be
strong, and we do not expect to see major changes in time in the
emission from known objects. 
On the other hand, any detection of major changes in the shape or polarization properties of a BSPWN could be a
clear indication for a strongly anisotropic energy injection and spin-axis misalignment.

Another important finding is related to the distribution of emitting
particles. 
In all of our models, if we consider a scaling proportional
to the local pressure, which is what one would naively expect for
particles uniformly accelerated at the wind termination shock and then
advected in the nebula, the head becomes much brighter that the
tail, even by a factor 10, and only in the most turbulent cases this
difference is less enhanced. 
There are indeed system like the Mouse nebula where a very bright head followed by a fainter tail is
observed, but in many others there is no evidence for the head to be
brighter than the tail \citep{Yusef-Zadeh:2005, Ng:2012}. 
This could be the signature of a diffuse acceleration process, or of peculiar injection conditions at
the termination shock. Interestingly there seems to be no appreciable difference between a uniform emissivity and an emissivity scaled
according to the strength of the currents. This means that it will be hard to distinguish these two possibility just based on the emission
pattern observed in known BSPWNe.

Another interesting aspect, related to polarization, is that in almost
all cases the direction of the polarization (the inferred direction of
the magnetic field) seems to be almost aligned with the tail (the only
exception is for the orthogonal case with high magnetization). This is
most likely due to polarization swing associated to the fast flow of
matter in the tail. In fact if one turns down relativistic beaming and
aberration, the structure of the polarization pattern changes, with a
tendency to be less aligned. 
Changes in the polarization direction are observed in BSPWNe, often associated with changes in brightness \citep{Ng_Gaensler+10a}. 
Our finding suggest that these could be due to rapid deceleration of the
flow in the tail, arising either as a consequence of internal shocks, or because of mass loading from the CD. The level of polarization is
instead a good indicator of turbulence, and generally scales with
magnetization. This could explain why we see system with polarization
as high as $60\%$ \citep{Ng:2012} and other as low as $10\%$ \citep{Yusef-Zadeh:2005}.

When magnetization is higher, and the dynamics in the tail is more
laminar, the emission appears to be less uniform, peaking in the head
region for the aligned case, $\phi_M=0$. When the inclination of the
pulsar spin-axis $\phi_M$ rises, the head starts to appear sub-luminous
($\phi_M=45^\circ$), with  some limb-brightening close to the CD for higher inclinations, surrounding a still quite sub-luminous area ($\phi_M=90^\circ$). The wings-to-core brightness also appear to increase with increasing $\phi_M$. 

The largest variety of morphologies in the maps are seen for the anisotropic wind cases, especially for the non aligned models ($\phi_M=45^\circ,\,\phi_M=90^\circ$). 
Of particular interest is the anisotropic case with $\phi_M=45^\circ$,
which shows very different emission patterns when changing the
observer's inclination angle, with an evident deformation from the
canonical cometary shape of the other models, resembling somehow the
 observed conical shape of the Mouse nebula (at least in the inner
 part). The separation of the head and tail direction of the
 polarization vector also resemble what observed in the Mouse nebula
 when the observer's viewing angle is not zero (especially for
 $\mathcal{X}=45^\circ$), with a clear component wrapped around the
 head and a general alignment with the nebular magnetic field in the
 tail.

We also computed a few maps considering a BSPWN where the pulsar
velocity does not lie on the plane of the sky. Given the limit on the
extent of the  domain of our simulations, we can only compute configurations where the PSR velocity
is inclined at most $45^\circ$ with respect to the plane of the sky
(for higher inclinations we cannot integrate along the line of sight in
the tail of the nebula without reaching the boundary of the numerical
model). The inclination does not change the wings-to-core behavior of
our maps, while it rises the intensity of the tail with
respect to the head. 
This can be easily understood in terms of integration along the line of sight. 
Given the cometary shape the inclination does not increase much the length of integration in the head, while it
rises its value in the tail by an amount that is roughly inversely
proportional the the cosine of the inclination angle. 
At $\mathcal{X}=45^\circ$ the tail is about twice brighter than in the orthogonal case shown in
Sect.~\ref{sec:analysis}, while the head has approximatively the same brightness. 

\section*{Acknowledgements}
We acknowledge the ``Accordo Quadro INAF-CINECA (2017-2019)''  for the availability of high performance computing resources and support. Simulations have been performed as part of the class-A project ``Three-dimensional relativistic simulations of bow shock nebulae'' (PI B. Olmi). 
The authors acknowledge financial support from the ``Accordo Attuativo ASI-INAF n. 2017-14-H.0 Progetto: \textit{on the escape of cosmic rays and their impact on the background plasma}'' and from the INFN Teongrav collaboration.
The authors wish to thank the referee, O. Toropina, for her timely and positive feedback on this work.
B. Olmi also acknowledges Andrea Mignone, from the PLUTO team, for fundamental support.

\footnotesize{
\bibliographystyle{mn2e}
\bibliography{olmi.bib}

\begin{thebibliography}{79}
\expandafter\ifx\csname natexlab\endcsname\relax\def\natexlab#1{#1}\fi

\bibitem[{{Abeysekara}(2017)}]{Abeysekara:2017}
{Abeysekara} A.~U. e.~a., 2017, Science, 358, 911

\bibitem[{{Amato} \& {Blasi}(2018)}]{Amato:2018}
{Amato} E., {Blasi} P., 2018, Advances in Space Research, 62, 2731

\bibitem[{{Arzoumanian}, {Chernoff} \& {Cordes}(2002){Arzoumanian}, {Chernoff},
  \& {Cordes}}]{Arzoumanian:2002}
{Arzoumanian} Z., {Chernoff} D.~F., {Cordes} J.~M., 2002, \apj, 568, 289

\bibitem[{{Arzoumanian} {et~al}\mbox{.}(2004){Arzoumanian}, {Cordes}, {Van
  Buren}, {Corcoran}, {Safi-Harb}, \& {Petre}}]{Arzoumanian_Cordes+04a}
{Arzoumanian} Z., {Cordes} J., {Van Buren} D., {Corcoran} M., {Safi-Harb} S.,
  {Petre} R., 2004, in Bulletin of the American Astronomical Society, Vol.~36,
  AAS/High Energy Astrophysics Division \#8, p. 951

\bibitem[{{Bandiera}(1993)}]{Bandiera93a}
{Bandiera} R., 1993, \aap, 276, 648

\bibitem[{{Barkov}, {Lyutikov} \& {Khangulyan}(2019){Barkov}, {Lyutikov}, \&
  {Khangulyan}}]{Barkov:2019}
{Barkov} M.~V., {Lyutikov} M., {Khangulyan} D., 2019, \mnras, 484, 4760

\bibitem[{{Bell} {et~al}\mbox{.}(1995){Bell}, {Bailes}, {Manchester},
  {Weisberg}, \& {Lyne}}]{Bell_Bailes+95a}
{Bell} J.~F., {Bailes} M., {Manchester} R.~N., {Weisberg} J.~M., {Lyne} A.~G.,
  1995, \apjl, 440, L81

\bibitem[{{Blasi} \& {Amato}(2011)}]{Blasi:2011}
{Blasi} P., {Amato} E., 2011, Astrophysics and Space Science Proceedings, 21,
  624

\bibitem[{{Bogovalov} \& {Tsinganos}(2001)}]{Bogovalov:2001}
{Bogovalov} S., {Tsinganos} K., 2001, Astronomical and Astrophysical
  Transactions, 20, 303

\bibitem[{{Brownsberger} \& {Romani}(2014)}]{Brownsberger:2014}
{Brownsberger} S., {Romani} R.~W., 2014, \apj, 784, 154

\bibitem[{{Bucciantini}(2002)}]{Bucciantini:2002}
{Bucciantini} N., 2002, \aap, 387, 1066

\bibitem[{{Bucciantini}(2018{\natexlab{a}})}]{Bucciantini:2018}
---, 2018{\natexlab{a}}, \mnras, 478, 2074

\bibitem[{{Bucciantini}(2018{\natexlab{b}})}]{Bucciantini:2018b}
---, 2018{\natexlab{b}}, \mnras, 480, 5419

\bibitem[{{Bucciantini}, {Amato} \& {Del Zanna}(2005){Bucciantini}, {Amato}, \&
  {Del Zanna}}]{Bucciantini:2005}
{Bucciantini} N., {Amato} E., {Del Zanna} L., 2005, \aap, 434, 189

\bibitem[{{Bucciantini} \& {Bandiera}(2001)}]{Bucciantini:2001}
{Bucciantini} N., {Bandiera} R., 2001, \aap, 375, 1032

\bibitem[{{Bucciantini} \& {Olmi}(2018)}]{Bucciantini:2018a}
{Bucciantini} N., {Olmi} B., 2018, \mnras, 475, 822

\bibitem[{{Caraveo} {et~al}\mbox{.}(2003){Caraveo}, {Bignami}, {De Luca},
  {Mereghetti}, {Pellizzoni}, {Mignani}, {Tur}, \& {Becker}}]{Caraveo:2003}
{Caraveo} P.~A., {Bignami} G.~F., {De Luca} A., {Mereghetti} S., {Pellizzoni}
  A., {Mignani} R., {Tur} A., {Becker} W., 2003, Science, 301, 1345

\bibitem[{{Cerutti} {et~al}\mbox{.}(2014){Cerutti}, {Werner}, {Uzdensky}, \&
  {Begelman}}]{Cerutti:2014}
{Cerutti} B., {Werner} G.~R., {Uzdensky} D.~A., {Begelman} M.~C., 2014, Physics
  of Plasmas, 21, 056501

\bibitem[{{Chatterjee} {et~al}\mbox{.}(2005){Chatterjee}, {Gaensler},
  {Vigelius}, {Cordes}, {Arzoumanian}, {Stappers}, {Ghavamian}, \&
  {Melatos}}]{Chatterjee:2005}
{Chatterjee} S., {Gaensler} B.~M., {Vigelius} M., {Cordes} J.~M., {Arzoumanian}
  Z., {Stappers} B., {Ghavamian} P., {Melatos} A., 2005, in Bulletin of the
  American Astronomical Society, Vol.~37, American Astronomical Society Meeting
  Abstracts, p. 1470

\bibitem[{{Chevalier}, {Kirshner} \& {Raymond}(1980){Chevalier}, {Kirshner}, \&
  {Raymond}}]{Chevalier_Kirshner+80a}
{Chevalier} R.~A., {Kirshner} R.~P., {Raymond} J.~C., 1980, \apj, 235, 186

\bibitem[{Childs {et~al}\mbox{.}(2012)Childs, Brugger, Whitlock, Meredith,
  Ahern, Pugmire, Biagas, Miller, Harrison, Weber, Krishnan, Fogal, Sanderson,
  Garth, Bethel, Camp, R\"{u}bel, Durant, Favre, \& Navr\'{a}til}]{Childs:2012}
Childs H. {et~al.}, 2012, in {High Performance Visualization--Enabling
  Extreme-Scale Scientific Insight}, Lawrence Berkeley National Laboratory, pp.
  357--372

\bibitem[{{Cioffi}, {McKee} \& {Bertschinger}(1988){Cioffi}, {McKee}, \&
  {Bertschinger}}]{Cioffi_McKee+88a}
{Cioffi} D.~F., {McKee} C.~F., {Bertschinger} E., 1988, \apj, 334, 252

\bibitem[{{Cordes} \& {Chernoff}(1998)}]{Cordes_Chernoff98a}
{Cordes} J.~M., {Chernoff} D.~F., 1998, \apj, 505, 315

\bibitem[{{Cordes}, {Romani} \& {Lundgren}(1993){Cordes}, {Romani}, \&
  {Lundgren}}]{Cordes:93}
{Cordes} J.~M., {Romani} R.~W., {Lundgren} S.~C., 1993, \nat, 362, 133

\bibitem[{{De Luca} {et~al}\mbox{.}(2011){De Luca}, {Marelli}, {Mignani},
  {Caraveo}, {Hummel}, {Collins}, {Shearer}, {Saz Parkinson}, {Belfiore}, \&
  {Bignami}}]{De-Luca_Marelli+11a}
{De Luca} A. {et~al.}, 2011, \apj, 733, 104

\bibitem[{{Del Zanna} {et~al}\mbox{.}(2006){Del Zanna}, {Volpi}, {Amato}, \&
  {Bucciantini}}]{Del-Zanna:2006}
{Del Zanna} L., {Volpi} D., {Amato} E., {Bucciantini} N., 2006, \aap, 453, 621

\bibitem[{{Dolch} {et~al}\mbox{.}(2016){Dolch}, {Chatterjee}, {Clemens},
  {Cordes}, {Cashmen}, \& {Taylor}}]{Dolch:2016}
{Dolch} T., {Chatterjee} S., {Clemens} D.~P., {Cordes} J.~M., {Cashmen} L.~R.,
  {Taylor} B.~W., 2016, Journal of Astronomy and Space Sciences, 33, 167

\bibitem[{{Gaensler}(2005)}]{Gaensler05a}
{Gaensler} B.~M., 2005, Advances in Space Research, 35, 1116

\bibitem[{{Gaensler} \& {Slane}(2006)}]{Gaensler:2006}
{Gaensler} B.~M., {Slane} P.~O., 2006, \araa, 44, 17

\bibitem[{{Gaensler} {et~al}\mbox{.}(2004){Gaensler}, {van der Swaluw},
  {Camilo}, {Kaspi}, {Baganoff}, {Yusef-Zadeh}, \&
  {Manchester}}]{Gaensler:2004}
{Gaensler} B.~M., {van der Swaluw} E., {Camilo} F., {Kaspi} V.~M., {Baganoff}
  F.~K., {Yusef-Zadeh} F., {Manchester} R.~N., 2004, \apj, 616, 383

\bibitem[{{Ghavamian} {et~al}\mbox{.}(2001){Ghavamian}, {Raymond}, {Smith}, \&
  {Hartigan}}]{Ghavamian_Raymond+01a}
{Ghavamian} P., {Raymond} J., {Smith} R.~C., {Hartigan} P., 2001, \apj, 547,
  995

\bibitem[{{Hales} {et~al}\mbox{.}(2009){Hales}, {Gaensler}, {Chatterjee}, {van
  der Swaluw}, \& {Camilo}}]{Hales_Gaensler+09a}
{Hales} C.~A., {Gaensler} B.~M., {Chatterjee} S., {van der Swaluw} E., {Camilo}
  F., 2009, \apj, 706, 1316

\bibitem[{{Hester}, {Raymond} \& {Blair}(1994){Hester}, {Raymond}, \&
  {Blair}}]{Hester_Raymond+94a}
{Hester} J.~J., {Raymond} J.~C., {Blair} W.~P., 1994, \apj, 420, 721

\bibitem[{{Hui} \& {Becker}(2007)}]{Hui_Becker07a}
{Hui} C.~Y., {Becker} W., 2007, \aap, 467, 1209

\bibitem[{{Jakobsen} {et~al}\mbox{.}(2014){Jakobsen}, {Tomsick}, {Watson},
  {Gotthelf}, \& {Kaspi}}]{Jakobsen_Tomsick+14a}
{Jakobsen} S.~J., {Tomsick} J.~A., {Watson} D., {Gotthelf} E.~V., {Kaspi}
  V.~M., 2014, \apj, 787, 129

\bibitem[{{Johnson} \& {Wang}(2010)}]{Johnson:2010}
{Johnson} S.~P., {Wang} Q.~D., 2010, \mnras, 408, 1216

\bibitem[{{Jones}, {Stappers} \& {Gaensler}(2002){Jones}, {Stappers}, \&
  {Gaensler}}]{Jones_Stappers+02a}
{Jones} D.~H., {Stappers} B.~W., {Gaensler} B.~M., 2002, \aap, 389, L1

\bibitem[{{Kargaltsev} {et~al}\mbox{.}(2008){Kargaltsev}, {Misanovic},
  {Pavlov}, {Wong}, \& {Garmire}}]{Kargaltsev_Misanovic+08a}
{Kargaltsev} O., {Misanovic} Z., {Pavlov} G.~G., {Wong} J.~A., {Garmire} G.~P.,
  2008, \apj, 684, 542

\bibitem[{{Kargaltsev}, {Pavlov} \& {Durant}(2012){Kargaltsev}, {Pavlov}, \&
  {Durant}}]{Kargaltsev:2012}
{Kargaltsev} O., {Pavlov} G.~G., {Durant} M., 2012, in Astronomical Society of
  the Pacific Conference Series, Vol. 466, Electromagnetic Radiation from
  Pulsars and Magnetars, {Lewandowski} W., {Maron} O., {Kijak} J., eds., p. 167

\bibitem[{{Kargaltsev} {et~al}\mbox{.}(2017){Kargaltsev}, {Pavlov}, {Klingler},
  \& {Rangelov}}]{Kargaltsev:2017}
{Kargaltsev} O., {Pavlov} G.~G., {Klingler} N., {Rangelov} B., 2017, Journal of
  Plasma Physics, 83, 635830501

\bibitem[{{Klingler} {et~al}\mbox{.}(2018){Klingler}, {Kargaltsev}, {Pavlov},
  {Ng}, {Beniamini}, \& {Volkov}}]{Klingler:2018}
{Klingler} N., {Kargaltsev} O., {Pavlov} G.~G., {Ng} C.-Y., {Beniamini} P.,
  {Volkov} I., 2018, \apj, 861, 5

\bibitem[{{Klingler} {et~al}\mbox{.}(2016{\natexlab{a}}){Klingler},
  {Kargaltsev}, {Rangelov}, {Pavlov}, {Posselt}, \& {Ng}}]{Klingler:2016}
{Klingler} N., {Kargaltsev} O., {Rangelov} B., {Pavlov} G.~G., {Posselt} B.,
  {Ng} C.-Y., 2016{\natexlab{a}}, \apj, 828, 70

\bibitem[{{Klingler} {et~al}\mbox{.}(2016{\natexlab{b}}){Klingler}, {Rangelov},
  {Kargaltsev}, {Pavlov}, {Romani}, {Posselt}, {Slane}, {Temim}, {Ng},
  {Bucciantini}, {Bykov}, {Swartz}, \& {Buehler}}]{Klingler_Rangelov+16a}
{Klingler} N. {et~al.}, 2016{\natexlab{b}}, \apj, 833, 253

\bibitem[{{Kulkarni} \& {Hester}(1988)}]{Kulkarni_Hester88a}
{Kulkarni} S.~R., {Hester} J.~J., 1988, \nat, 335, 801

\bibitem[{{Leahy}, {Green} \& {Tian}(2014){Leahy}, {Green}, \&
  {Tian}}]{Leahy_Green+14a}
{Leahy} D., {Green} K., {Tian} W., 2014, \mnras, 438, 1813

\bibitem[{{Li}, {Lu} \& {Li}(2005){Li}, {Lu}, \& {Li}}]{Li:2005}
{Li} X.~H., {Lu} F.~J., {Li} T.~P., 2005, \apj, 628, 931

\bibitem[{{Lyutikov} {et~al}\mbox{.}(2016){Lyutikov}, {Sironi}, {Komissarov},
  \& {Porth}}]{Lyutikov:2016}
{Lyutikov} M., {Sironi} L., {Komissarov} S., {Porth} O., 2016, arXiv e-prints,
  arXiv:1603.05731

\bibitem[{{Marelli} {et~al}\mbox{.}(2013){Marelli}, {De Luca}, {Salvetti},
  {Sartore}, {Sartori}, {Caraveo}, {Pizzolato}, {Saz Parkinson}, \&
  {Belfiore}}]{Marelli_De-Luca+13a}
{Marelli} M. {et~al.}, 2013, \apj, 765, 36

\bibitem[{{Marelli} {et~al}\mbox{.}(2019){Marelli}, {Tiengo}, {De Luca},
  {Mignani}, {Salvetti}, {Saz Parkinson}, \& {Lisini}}]{Marelli:2019}
{Marelli} M., {Tiengo} A., {De Luca} A., {Mignani} R.~P., {Salvetti} D., {Saz
  Parkinson} P.~M., {Lisini} G., 2019, \aap, 624, A53

\bibitem[{{Michel}(1973)}]{Michel:1973}
{Michel} F.~C., 1973, \apjl, 180, L133

\bibitem[{{Misanovic}, {Pavlov} \& {Garmire}(2008){Misanovic}, {Pavlov}, \&
  {Garmire}}]{Misanovic_Pavlov+08a}
{Misanovic} Z., {Pavlov} G.~G., {Garmire} G.~P., 2008, \apj, 685, 1129

\bibitem[{{Morlino}, {Lyutikov} \& {Vorster}(2015){Morlino}, {Lyutikov}, \&
  {Vorster}}]{Morlino:2015}
{Morlino} G., {Lyutikov} M., {Vorster} M., 2015, \mnras, 454, 3886

\bibitem[{{Ng} {et~al}\mbox{.}(2012){Ng}, {Bucciantini}, {Gaensler}, {Camilo},
  {Chatterjee}, \& {Bouchard}}]{Ng:2012}
{Ng} C.-Y., {Bucciantini} N., {Gaensler} B.~M., {Camilo} F., {Chatterjee} S.,
  {Bouchard} A., 2012, \apj, 746, 105

\bibitem[{{Ng} {et~al}\mbox{.}(2009){Ng}, {Camilo}, {Chatterjee}, {Gaensler},
  {Yusef-Zadeh}, {Hales}, {Johnston}, {Manchester}, {Kuiper}, \& {van der
  Swaluw}}]{Ng_Camilo+09a}
{Ng} C.~Y. {et~al.}, 2009, in Bulletin of the American Astronomical Society,
  Vol.~41, American Astronomical Society Meeting Abstracts \#213, p. 307

\bibitem[{{Ng} {et~al}\mbox{.}(2010){Ng}, {Gaensler}, {Chatterjee}, \&
  {Johnston}}]{Ng_Gaensler+10a}
{Ng} C.-Y., {Gaensler} B.~M., {Chatterjee} S., {Johnston} S., 2010, \apj, 712,
  596

\bibitem[{{Olmi} \& {Bucciantini}(2019)}]{Olmi:2019}
{Olmi} B., {Bucciantini} N., 2019, \mnras, 484, 5755

\bibitem[{{Olmi}, {Bucciantini} \& {Morlino}(2018){Olmi}, {Bucciantini}, \&
  {Morlino}}]{Olmi:2018}
{Olmi} B., {Bucciantini} N., {Morlino} G., 2018, \mnras, 481, 3394

\bibitem[{{Olmi} {et~al}\mbox{.}(2014){Olmi}, {Del Zanna}, {Amato}, {Bandiera},
  \& {Bucciantini}}]{Olmi:2014}
{Olmi} B., {Del Zanna} L., {Amato} E., {Bandiera} R., {Bucciantini} N., 2014,
  \mnras, 438, 1518

\bibitem[{{Pavan} {et~al}\mbox{.}(2014){Pavan}, {Bordas}, {P{\"u}hlhofer},
  {Filipovi{\'c}}, {De Horta}, {O'Brien}, {Balbo}, {Walter}, {Bozzo},
  {Ferrigno}, {Crawford}, \& {Stella}}]{Pavan_Bordas+14a}
{Pavan} L. {et~al.}, 2014, \aap, 562, A122

\bibitem[{{Pavlov}, {Bhattacharyya} \& {Zavlin}(2010){Pavlov}, {Bhattacharyya},
  \& {Zavlin}}]{Pavlov_Bhattacharyya+10a}
{Pavlov} G.~G., {Bhattacharyya} S., {Zavlin} V.~E., 2010, \apj, 715, 66

\bibitem[{{Pavlov}, {Sanwal} \& {Zavlin}(2006){Pavlov}, {Sanwal}, \&
  {Zavlin}}]{Pavlov:2006}
{Pavlov} G.~G., {Sanwal} D., {Zavlin} V.~E., 2006, \apj, 643, 1146

\bibitem[{{Posselt} {et~al}\mbox{.}(2017){Posselt}, {Pavlov}, {Slane},
  {Romani}, {Bucciantini}, {Bykov}, {Kargaltsev}, {Weisskopf}, \&
  {Ng}}]{Posselt_Pavlov+17a}
{Posselt} B. {et~al.}, 2017, \apj, 835, 66

\bibitem[{{Rangelov} {et~al}\mbox{.}(2016){Rangelov}, {Pavlov}, {Kargaltsev},
  {Durant}, {Bykov}, \& {Krassilchtchikov}}]{Rangelov_Pavlov+16a}
{Rangelov} B., {Pavlov} G.~G., {Kargaltsev} O., {Durant} M., {Bykov} A.~M.,
  {Krassilchtchikov} A., 2016, \apj, 831, 129

\bibitem[{{Romani}, {Slane} \& {Green}(2017){Romani}, {Slane}, \&
  {Green}}]{Romani_Slane+17a}
{Romani} R.~W., {Slane} P., {Green} A.~W., 2017, \apj, 851, 61

\bibitem[{{S{\'a}nchez-Cruces} {et~al}\mbox{.}(2018){S{\'a}nchez-Cruces},
  {Rosado}, {Fuentes-Carrera}, \& {Ambrocio-Cruz}}]{Sanchez-Cruces_Rosado+18a}
{S{\'a}nchez-Cruces} M., {Rosado} M., {Fuentes-Carrera} I., {Ambrocio-Cruz} P.,
  2018, \mnras, 473, 1705

\bibitem[{{Sartore} {et~al}\mbox{.}(2010){Sartore}, {Ripamonti}, {Treves}, \&
  {Turolla}}]{Sartore_Ripamonti+10a}
{Sartore} N., {Ripamonti} E., {Treves} A., {Turolla} R., 2010, \aap, 510, A23

\bibitem[{{Spitkovsky}(2006)}]{Spitkovsky:2006}
{Spitkovsky} A., 2006, \apjl, 648, L51

\bibitem[{{Truelove} \& {McKee}(1999)}]{Truelove_McKee99a}
{Truelove} J.~K., {McKee} C.~F., 1999, \apjs, 120, 299

\bibitem[{{Uzdensky}, {Cerutti} \& {Begelman}(2011){Uzdensky}, {Cerutti}, \&
  {Begelman}}]{Uzdensky_Cerutti+11a}
{Uzdensky} D.~A., {Cerutti} B., {Begelman} M.~C., 2011, \apjl, 737, L40

\bibitem[{{van der Swaluw} {et~al}\mbox{.}(2003){van der Swaluw}, {Achterberg},
  {Gallant}, {Downes}, \& {Keppens}}]{van-der-Swaluw:2003}
{van der Swaluw} E., {Achterberg} A., {Gallant} Y.~A., {Downes} T.~P.,
  {Keppens} R., 2003, \aap, 397, 913

\bibitem[{{van Kerkwijk} \& {Kulkarni}(2001)}]{van-Kerkwijk_Kulkarni01a}
{van Kerkwijk} M.~H., {Kulkarni} S.~R., 2001, \aap, 380, 221

\bibitem[{{Verbunt}, {Igoshev} \& {Cator}(2017){Verbunt}, {Igoshev}, \&
  {Cator}}]{Verbunt_Igoshev+17a}
{Verbunt} F., {Igoshev} A., {Cator} E., 2017, \aap, 608, A57

\bibitem[{{Vigelius} {et~al}\mbox{.}(2007){Vigelius}, {Melatos}, {Chatterjee},
  {Gaensler}, \& {Ghavamian}}]{Vigelius:2007}
{Vigelius} M., {Melatos} A., {Chatterjee} S., {Gaensler} B.~M., {Ghavamian} P.,
  2007, \mnras, 374, 793

\bibitem[{{Volpi} {et~al}\mbox{.}(2008){Volpi}, {Del Zanna}, {Amato}, \&
  {Bucciantini}}]{Volpi:2008}
{Volpi} D., {Del Zanna} L., {Amato} E., {Bucciantini} N., 2008, \aap, 485, 337

\bibitem[{{Wang} {et~al}\mbox{.}(2013){Wang}, {Kaplan}, {Slane}, {Morrell}, \&
  {Kaspi}}]{Wang_Kaplan+13a}
{Wang} Z., {Kaplan} D.~L., {Slane} P., {Morrell} N., {Kaspi} V.~M., 2013, \apj,
  769, 122

\bibitem[{{Wilkin}(1996)}]{Wilkin:1996}
{Wilkin} F.~P., 1996, \apjl, 459, L31

\bibitem[{{Wong} {et~al}\mbox{.}(2003){Wong}, {Cordes}, {Chatterjee},
  {Zweibel}, {Finley}, {Romani}, \& {Ulmer}}]{Wong:2003}
{Wong} D.~S., {Cordes} J.~M., {Chatterjee} S., {Zweibel} E.~G., {Finley} J.~P.,
  {Romani} R.~W., {Ulmer} M.~P., 2003, in IAU Symposium, Vol. 214, High Energy
  Processes and Phenomena in Astrophysics, {Li} X.~D., {Trimble} V., {Wang}
  Z.~R., eds., p. 135

\bibitem[{{Yusef-Zadeh} \& {Bally}(1987)}]{Yusef-Zadeh:1987}
{Yusef-Zadeh} F., {Bally} J., 1987, \nat, 330, 455

\bibitem[{{Yusef-Zadeh} \& {Gaensler}(2005)}]{Yusef-Zadeh:2005}
{Yusef-Zadeh} F., {Gaensler} B.~M., 2005, Advances in Space Research, 35, 1129

\end{thebibliography}
}

\end{document}